\documentclass[iop,revtex4]{emulateapj}
\newcommand{\myemail}{toshiki.saito@nao.ac.jp}
\usepackage[breaklinks,colorlinks,urlcolor=blue,citecolor=blue,linkcolor=blue]{hyperref}
\defcitealias{Saito15b}{S16a}

\shorttitle{CO SLED Imaging of NGC~1614 with ALMA}
\shortauthors{Toshiki Saito et al.}

\begin{document}

\title{Spatially resolved CO SLED of the Luminous Merger Remnant NGC~1614 with ALMA}

\author{
   Toshiki \textsc{Saito}\altaffilmark{1,2},
   Daisuke \textsc{Iono}\altaffilmark{2,3},
   Cong \textsc{K. Xu}\altaffilmark{4},
   Kazimierz \textsc{Sliwa}\altaffilmark{5},
   Junko \textsc{Ueda}\altaffilmark{6},
   Daniel \textsc{Espada}\altaffilmark{2,3},
   Hiroyuki \textsc{Kaneko}\altaffilmark{7},
   Sabine \textsc{K\"{o}nig}\altaffilmark{8},
   Kouichiro \textsc{Nakanishi}\altaffilmark{2,3},
   Minju \textsc{Lee}\altaffilmark{1,2},
   Min \textsc{S. Yun}\altaffilmark{9},
   Susanne \textsc{Aalto}\altaffilmark{8},
   John \textsc{E. Hibbard}\altaffilmark{10},
   Takuji \textsc{Yamashita}\altaffilmark{11},
   Kentaro \textsc{Motohara}\altaffilmark{12},
   and
   Ryohei \textsc{Kawabe}\altaffilmark{1,2,3}
}
\email{\myemail}
 \altaffiltext{1}{Department of Astronomy, The University of Tokyo, 7-3-1 Hongo, Bunkyo-ku, Tokyo 113-0033, Japan}
 \altaffiltext{2}{National Astronomical Observatory of Japan, 2-21-1 Osawa, Mitaka, Tokyo, 181-8588, Japan}
 \altaffiltext{3}{The Graduate University for Advanced Studies (SOKENDAI), 2-21-1 Osawa, Mitaka, Tokyo, 181-0015, Japan}
 \altaffiltext{4}{Infrared Processing and Analysis Center (IPAC), California Institute of Technology, 770 South Wilson Avenue, Pasadena, CA 91125, USA}
  \altaffiltext{5}{Max-Planck Institute for Astronomy, K\"{o}nigstuhl 17, D-69117 Heidelberg, Germany}
  \altaffiltext{6}{Harvard-Smithsonian Center for Astrophysics, 60 Garden Street, Cambridge, MA 02138, USA}
  \altaffiltext{7}{Nobeyama Radio Observatory, Minamimaki, Minamisaku, Nagano 384-1305, Japan}
  \altaffiltext{8}{Chalmers University of Technology, Department of Earth and Space Sciences, Onsala Space Observatory, 43992 Onsala, Sweden}
 \altaffiltext{9}{Department of Astronomy, University of Massachusetts, Amherst, MA 01003, USA}
 \altaffiltext{10}{National Radio Astronomy Observatory, 520 Edgemont Road, Charlottesville, VA, 22903, USA}
 \altaffiltext{11}{Institute of Space and Astronautical Science, Japan Aerospace Exploration Agency, 3-1-1 Yoshinodai, Sagamihara, Kanagawa 229-8510, Japan}
 \altaffiltext{12}{Institute of Astronomy, The University of Tokyo, 2-21-1 Osawa, Mitaka, Tokyo 181-0015, Japan}

\received{October 23, 2016}
\revised{December 8, 2016}
\accepted{December 21, 2016}

\begin{abstract}

We present high-resolution (1\farcs0) Atacama Large Millimeter/submillimeter Array (ALMA) observations of CO~(1--0) and CO~(2--1) rotational transitions toward the nearby IR-luminous merger NGC~1614 supplemented with ALMA archival data of CO~(3--2), and CO~(6--5) transitions.
The CO~(6--5) emission arises from the starburst ring (central 590~pc in radius), while the lower-$J$ CO lines are distributed over the outer disk ($\sim$ 3.3~kpc in radius).
Radiative transfer and photon dominated region (PDR) modeling reveal that the starburst ring has a single warmer gas component with more intense far-ultraviolet radiation field ($n_{\rm{H_2}}$ $\sim$ 10$^{4.6}$ cm$^{-3}$, $T_{\rm{kin}}$ $\sim$ 42~K, and $G_{\rm{0}}$ $\sim$ 10$^{2.7}$) relative to the outer disk ($n_{\rm{H_2}}$ $\sim$ 10$^{5.1}$ cm$^{-3}$, $T_{\rm{kin}}$ $\sim$ 22~K, and $G_{\rm{0}}$ $\sim$ 10$^{0.9}$).
A two-phase molecular interstellar medium with a warm and cold ($>$ 70~K and $\sim$ 19~K) component is also an applicable model for the starburst ring.
A possible source for heating the warm gas component is mechanical heating due to stellar feedback rather than PDR.
Furthermore, we find evidence for non-circular motions along the north-south optical bar in the lower-$J$ CO images, suggesting a cold gas inflow.
We suggest that star formation in the starburst ring is sustained by the bar-driven cold gas inflow, and starburst activities radiatively and mechanically power the CO excitation.
The absence of a bright active galactic nucleus can be explained by a scenario that cold gas accumulating on the starburst ring is exhausted as the fuel for star formation, or is launched as an outflow before being able to feed to the nucleus.

\end{abstract}


\section{INTRODUCTION}

Recent single-dish and {\it Herschel} spectroscopic observations successfully detected bright high-$J$ CO emission \citep[$J$ = 4 -- 3 up to 30 -- 29;][]{Panuzzo10,van_der_Werf10,Rangwala11,Hailey-Dunsheath12,Kamenetzky12,Papadopoulos12,Rosenberg12,Spinoglio12,Meijerink13,Pellegrini13,Pereira-Santaella13,Rigopoulou13,Greve14,Kamenetzky14, Lu14,Papadopoulos14,Pereira-Santaella14,Schirm14,Topal14, Falstad15,Liu15,Mashian15, Rosenberg15,Wu15,Kamenetzky16,Pearson16} from nearby starburst galaxies and (ultra-)luminous infrared galaxies (U/LIRGs), as well as lower-$J$ CO lines \citep[e.g.,][]{Yao03,Narayanan05,Iono09,Leech10,Mao10,Papadopoulos12,Wilson12,Michiyama16}.
The observational evidence of extreme CO excitation in U/LIRGs is explained by a combination of several heating models of interstellar medium (ISM) which consists of cosmic-ray dominated regions (CRDRs), photon dominated regions (PDRs), X-ray dominated regions (XDRs), and/or mechanically dominated regions (MDRs), because coarse single-dish beams ($\lesssim$ 10~kpc) are not possible to distinguish many molecular conditions \citep{Papadopoulos10, van_der_Werf10, Bayet11, Aalto13, Meijerink13}.
High-resolution imaging of various CO lines is, thus, an important way to investigate the multiple phases of the molecular ISM and physical processes involved in obscured central activities of galaxies.

\begin{figure*}
\begin{center}
\includegraphics[width=18cm]{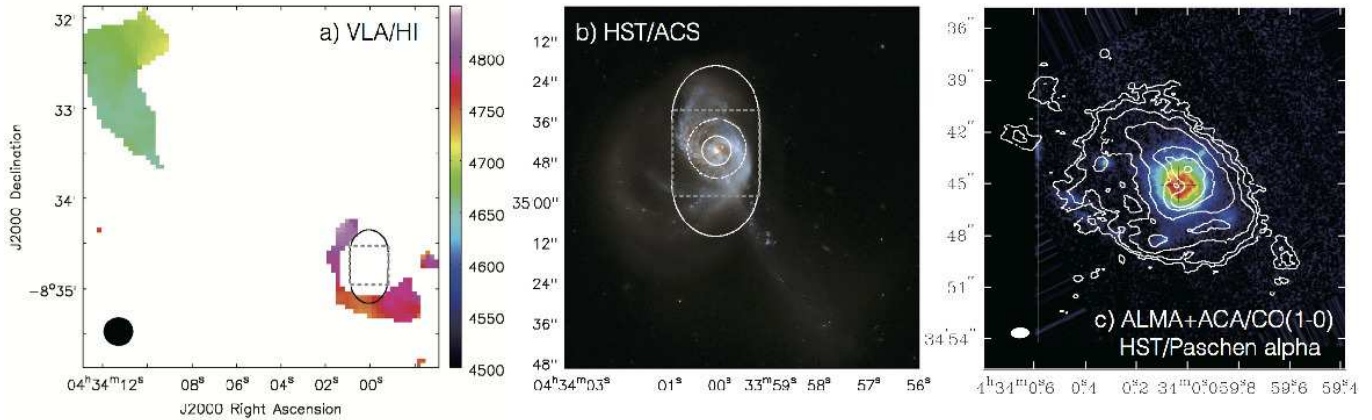}
\caption{(a) HI image of NGC~1614 obtained by the Very Large Array \citep{Hibbard&Yun96}.  The velocity field in color scale ranges from 4500 km s$^{-1}$ to 4850 km s$^{-1}$.  The black ellipse shows the field of view of the Band~6 observation.  The dashed square corresponds to the imaging area of Figure~\ref{figure_1}c.  (b) HST/ACS image of NGC~1614 [Credit: NASA, ESA, the Hubble Heritage Team (STScI/AURA)-ESA/Hubble Collaboration and A. Evans (University of Virginia, Charlottesville/NRAO/Stony Brook University)].  The white ellipse shows the field of view of the Band~6 observation. The dashed and solid circles show the field of view of the Band~7 and Band~9 observations, respectively.  The field of view of the Band~3 observation is three times larger than that of the Band~7 observation.  (c) ACA-combined CO~(1--0) integrated intensity contour overlaid on Paschen $\alpha$ image \citep{Alonso-Herrero01}.  The contours are 8.87 $\times$ (0.03, 0.06, 0.12, 0.24, 0.48, 0.64, and 0.96) Jy beam$^{-1}$ km s$^{-1}$.  The black cross indicates the nucleus which is detected in Pa $\alpha$ and the radio continuum emission \citep{Olsson10,Herrero-Illana14}.  The imaging area is same as CO images shown in Figure~\ref{figure_2}.
}
\label{figure_1}
\end{center}
\end{figure*}

Interferometric studies of CO $J$=6--5 transition (hereafter CO~(6--5)) in nearby LIRGs showed that the distribution of the CO~(6--5) emission is compact compared to the extended lower-$J$ CO emission \citep{Matsushita09, Sliwa13, Sliwa14, Xu14, Rangwala15, Xu15, Zhao16}, suggesting that CO~(6--5) probes warmer and denser gas directly related to the nuclear activities.
However, a direct comparison among distributions of multiple CO transitions is not straightforward as we are limited in angular resolution for the low-$J$ lines, sensitivities and $uv$-coverages vary, and the high-$J$ lines require good weather making observations scarce from the ground.
In this paper, we present high-resolution, high-sensitivity, and $uv$-matched ALMA observations of the nearby IR-bright galaxy NGC~1614 in order to understand the distribution and kinematics of multiple CO lines and their excitation conditions in the nuclear region.
We investigate lower-$J$ CO lines up to $J_{\rm{upp}}$ = 6, which are thought to be mainly excited by star-forming activities.

NGC~1614 is a nearby LIRG \citep[$L_{\rm{IR}}$ = 10$^{11.65}$ $L_{\odot}$;][]{Armus09} at a distance of 67.8 Mpc (1\arcsec = 330~pc).
The total star formation rate (SFR) is $\sim$ 75~$M_{\odot}$ yr$^{-1}$ based on the measurement of extinction-corrected Paschen $\alpha$ emission \citep{Tateuchi15}.
Merging signatures such as a tidal tail are clearly seen in the HI \citep[Figure \ref{figure_1}a;][]{Hibbard&Yun96} and optical images \citep[Figure \ref{figure_1}b;][]{Alonso-Herrero01}.
Numerical simulations performed by \citet{Vaisanen12} suggested that the system is a minor merger with the mass ratio of 1:3 -- 1:5 (i.e., nearly major merger).
A strong HI absorption is detected toward the central region of NGC~1614 surrounded by an arc-like structure in HI emission \citep{Hibbard&Yun96}.
Visually, NGC~1614 has been classified as a merger remnant because of no nearby companion \citep{Rothberg&Joseph04}, and is one of the remnants with a rotating molecular gas disk which possibly evolves into a late-type galaxy \citep{Ueda14}.
HCN~(4--3) and HCO$^+$~(4--3) observations show that NGC~1614 is consistent with a starburst-dominated galaxy without strong active galactic nuclei (AGN) signatures \citep{Costagliola11, Imanishi&Nakanishi13}.
This is consistent with the radio, mid-IR, and X-ray observations \citep{Herrero-Illana14, Pereira-Santaella15}.
The global CO spectral line energy distribution (SLED) is characterized by a steep decline towards higher-$J$ transitions \citep{Rosenberg15}.
High-resolution ($\sim$ 0\farcs25 $\sim$ 83~pc) ALMA Band~9 observations revealed that the CO~(6--5) distribution coincides with a starburst ring (r $<$ 350~pc) detected in Pa $\alpha$ \citep[Figure~\ref{figure_1}c;][]{Alonso-Herrero01} and 8.4~GHz radio continuum emission \citep{Xu15}, while the CO~(2--1) and CO~(3--2) emission are more extended \citep{Wilson08, Konig13, Sliwa14}.
Recent high-sensitivity CO~(2--1) observations revealed the presence of a high-velocity wing, which might be associated with a nuclear molecular outflow \citep{Garcia-burillo15}, whereas deep CO~(1--0) imaging found that extended molecular gas components associated with the southern tidal tail and a dust lane \citep{Konig16}.
Continuum emission from 4.81~GHz to 691~GHz is confined within the central 1\farcs7 radius (= 560~pc) and the radio-to-FIR spectral energy distribution (SED) can be explained by star-forming activities (\citealt{Saito15b}, hereafter \citetalias{Saito15b}).

This Paper is organized as follows: the observations and data reduction are summarized in Section~\ref{obs}.
All CO maps, position-velocity diagrams, and line intensity ratios are presented in Section~\ref{results}.
In Section~\ref{model}, we describe modeling procedures of the spatially-resolved CO SLED using three different models.
We provide a discussion of our modeling results in Section~\ref{discuss}, including heating source, CO-to-H$_2$ conversion factor, a possible two-phase ISM model and its heating source, cold molecular gas kinematics (inflow and outflow), and overall picture of molecular gas in NGC~1614.
Finally, we summarize our main findings in Section~\ref{conclusion}.
We have adopted H$_0$ = 70 km s$^{-1}$ Mpc$^{-1}$, $\Omega_m$ = 0.3, $\Omega_{\Lambda}$ = 0.7 throughout this Paper.

\section{ALMA OBSERVATIONS AND DATA REDUCTION} \label{obs}
\subsection{Band~3 and Band~6 Data: 12~m Array} \label{12}
The CO~(1--0) and CO~(2--1) observations toward NGC~1614 were carried out for our ALMA Cycle~2 program (ID: 2013.1.01172.S) using thirty-five 12~m antennas.
The Band~3 and Band~6 receivers were tuned to the CO~(1--0) and CO~(2--1) emission lines in the upper sideband, respectively.
The Band~3 data (single pointing), with a projected baseline length ($L_{\rm{baseline}}$) of 28 - 1060~m (C34-6 configuration), were obtained on 2014 August 30 (on-source time of $T_{\rm{integ}}$ = 16.9~min.), and the Band~6 data (3-point mosaic) with $L_{\rm{baseline}}$ = 15 - 349~m (C34-2/1 configuration) were obtained on 2014 December 8 ($T_{\rm{integ}}$ = 7.3~min per pointing).
The correlator was configured to have four spectral windows (SPW), two of which were set to each sideband, each of the SPWs with 1.875~GHz bandwidth and 1.129 MHz resolution ($\sim$ 3.0~km s$^{-1}$ and 1.4~km s$^{-1}$ for Band~3 and Band~6, respectively).
Both Bands used a strong quasar J0423-0120 for bandpass and phase calibration.
J0423-0120 (= J0423-013) was also the flux calibrator in Band~3, while Uranus was that in Band~6.

\subsection{Ancillary Data}
\subsubsection{Band~7 and Band~9 Data: 12~m Array}
NGC 1614 was observed during Cycle 0 using ALMA in Bands 7 and 9 (ID: 2011.0.00182.S and 2011.0.00768.S), and the data were originally published in \citet{Sliwa14} and \citet{Xu15}.
We obtained the calibrated archival visibility data of the CO~(3--2) and the CO~(6--5) from the ALMA archive.
Detailed information of the Cycle 0 and 2 data are shown in Table~\ref{table_obs} (see also \citetalias{Saito15b}).

\subsubsection{Band~3 Data: Atacama Compact Array and 12~m Array} \label{ACA}
The CO~(1--0) observations toward NGC~1614 were also carried out during Cycle~2 \citep[ID: 2013.1.00991.S;][]{Konig16} using both ten 7~m antennas of ACA \citep{Iguchi09} and thirty-one 12~m antennas of the 12~m array.
These ancillary data are used to increase the sensitivity and recover extended emission that we ignore due to the $uv$-clipping (Section~\ref{12}).
Thus, we use those data to study the global gas kinematics of NGC~1614.
we retrieved those data from the ALMA archive, and used one SPW which was tuned to the CO~(1--0) line.
Detailed description of this project is found in \citet{Konig16}, and we list some parameters in Table~\ref{table_data}.

\subsection{Reduction Procedure}
\subsubsection{12~m Array Data}
We performed calibration and imaging using {\tt CASA} \citep[version 3.4, 4.1.0, and 4.2.2 for Band~7, 9, and 3/6, respectively;][]{McMullin07}.
Detailed procedure for data processing is summarized in \citetalias{Saito15b}.
Before imaging, we flagged the $uv$ range below 45 k$\lambda$, so that the maximum recoverable scale \citep[MRS;][]{Lundgen13} of each ALMA observation is consistent.
The minimum $uv$ distance is determined by the configuration of the CO~(6--5) observation.
Since the truncated $uv$ range at 45 k$\lambda$ corresponds to the MRS of $\sim$ 4\farcs6 ($\simeq$ 1.5~kpc), we ignore the missing flux effect on structures smaller than 4\farcs6 throughout this Paper.
All of the images are convolved to the same resolution (1\farcs0 $\times$ 1\farcs0).
We made all maps in this Paper without the primary beam correction, although flux measurements were done after that.
The systematic errors on the absolute flux calibration using a solar system object are estimated to be 5\%, 10\%, 10\%, and 15\% for both sidebands in Band~3, Band~6, Band~7, and Band~9, respectively \citep{Lundgen13}.
The Band~3 observation has a quasar as the flux calibrator, so we estimate the absolute systematic error on the flux calibration for our Band~3 data by using the ALMA Calibrator Source Catalogue\footnote{https://almascience.nrao.edu/sc/}.
The flux uncertainty of J0423-013 on 2014 August 30 is 1.9-3.7\% at 103.5~GHz.
Therefore, we adopt the flux uncertainty of 5\% to be conservative.
Since the recovered flux of each CO transition compared to single-dish observations \citep{Sanders91, Albrecht07, Wilson08, Xu15} is very low (27 - 59~\%), NGC~1614 has large amounts of ambient molecular gas structures larger than 1.5~kpc even in the CO~(6--5).
The detailed data properties and line information are listed in Table~\ref{table_data}.

\subsubsection{ACA Data and Combine with 12~m Array Data}
We used {\tt CASA} (version 4.2.1 and 4.2.2) to process the data.
In order to combine the CO~(1--0) data shown in Section~\ref{ACA} with the data described in Section~\ref{12}, we used {\tt statwt}\footnote{https://casaguides.nrao.edu/index.php/DataWeightsAnd\\Combination} task to recalculate visibility weights for all the data.
The resultant $L_{\rm{baseline}}$ is 7.3 - 1060~m.
This corresponds to the MRS of 17\farcs6 ($\sim$ 5.8~kpc).
We constructed a data cube with the velocity resolution of 20~km s$^{-1}$.
The flux calibrator for the ACA observation is a quasar J0501-0159.
Using the ALMA Calibrator Source Catalogue\footnotemark[13], the flux uncertainty of J0501-0159 on 2015 June 30 is estimated to be $\sim$ 12\% at 97.4~GHz.
Therefore, we adopt the flux uncertainty of 12\% when we use the ACA-combined CO~(1--0) data.
The achieved rms noise level is 1.4~mJy beam$^{-1}$ with the synthesized beam size of 1\farcs02 $\times$ 0\farcs57 (position angle = $-$88\degr).
Comparing with the CO~(1--0) flux obtained by NRAO 12m \citep{Sanders91}, the recovered flux is 113 $\pm$ 26~\% indicating that we recovered most (or probably all) of the CO~(1--0) emission.
The total H$_2$ mass is (4.1 $\pm$ 0.5) $\times$ 10$^9$~$M_{\odot}$ assuming 1/5 of the Milky Way CO-to-H$_2$ conversion factor \citep[$\alpha_{\rm{MW}}$ = 4.3 $M_{\odot}$/(K km s$^{-1}$ pc$^2$)$^{-1}$;][]{Downes&Solomon98,Bolatto13}.
When comparing with ACA-only CO~(1--0) flux measured by \citet{Konig16} ($\sim$ 241 Jy km s$^{-1}$), the recovered flux is $\sim$ 123~\%.
This difference might be due to the differences of mask shape to measure the CO flux in R.A., Decl., and velocity axes (i.e., low surface brightness, high velocity wings).
Detailed information of the ACA and the 12~m array data is shown in Table~\ref{table_obs}, and the line information are listed in Table~\ref{table_data}.

\begin{figure*}[!t]
\begin{center}
\includegraphics[width=16cm]{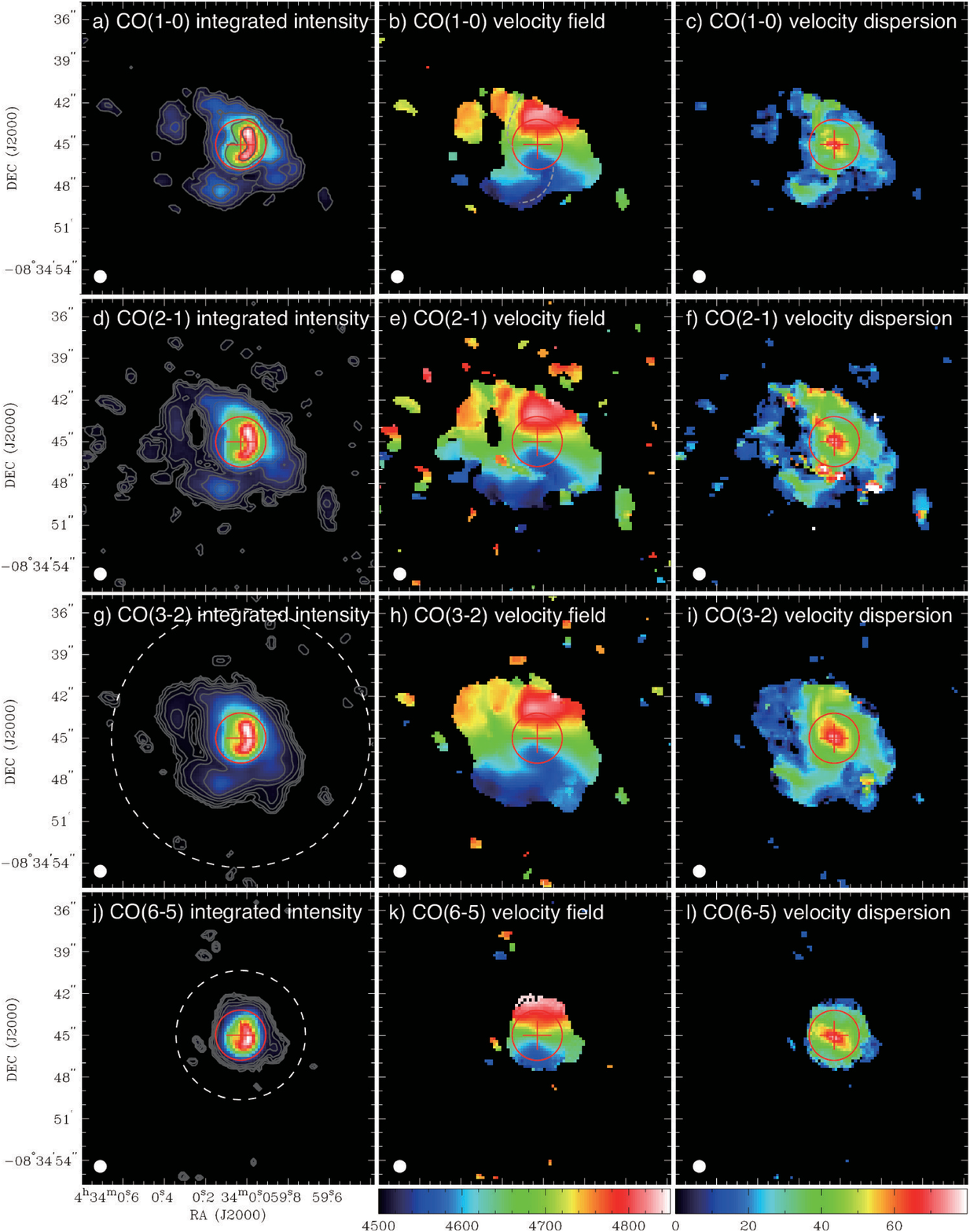}
\caption{(a) $uv$-clipped CO~(1--0) integrated intensity image of NGC~1614.  The contours are 0.36, 0.72, 1.44, 2.88, 5.76, and 11.52 Jy beam$^{-1}$ km s$^{-1}$.  The red cross indicates the nucleus which is detected in Pa $\alpha$ and radio continuum emission \citep{Olsson10,Herrero-Illana14}.  The red circle indicates the approximate outer edge of the starburst ring (\citealt{Konig13,Herrero-Illana14, Xu15}; \citetalias{Saito15b}).  (b) $uv$-clipped CO~(1--0) velocity field image.  The velocity field in color scale ranges from 4500 km s$^{-1}$ to 4850 km s$^{-1}$.  The dashed grey lines show the positions of non-circular motion.  (c) $uv$-clipped CO~(1--0) velocity dispersion image.  The velocity dispersion in color scale ranges from 0 km s$^{-1}$ to 80 km s$^{-1}$.  (d) The same as (a) but for CO~(2--1).  The contours are 1.0, 2.0, 4.0, 8.0, 16.0, and 32.0 Jy beam$^{-1}$ km s$^{-1}$.  (e/f) The same as (b/c), respectively, but for CO~(2--1).  (g) The same as (a) but for CO~(3--2).  The contours are 0.72, 1.44, 2.88, 5.76, 11.52, and 23.04 Jy beam$^{-1}$ km s$^{-1}$.  (h/i) The same as (b/c), respectively, but for CO~(3--2).  (j) The same as (a) but for CO~(6--5).  The contours are 2.4, 4.8, 9.6, 19.2, 38.4, and 76.8 Jy beam$^{-1}$ km s$^{-1}$.  (k/l) The same as (b/c), respectively, but for CO~(6--5).  The fields of view of the CO~(3--2) and CO~(6--5) data are shown as white dashed circles.
}
\label{figure_2}
\end{center}
\end{figure*}

\begin{figure*}[!t]
\begin{center}
\includegraphics[width=15cm]{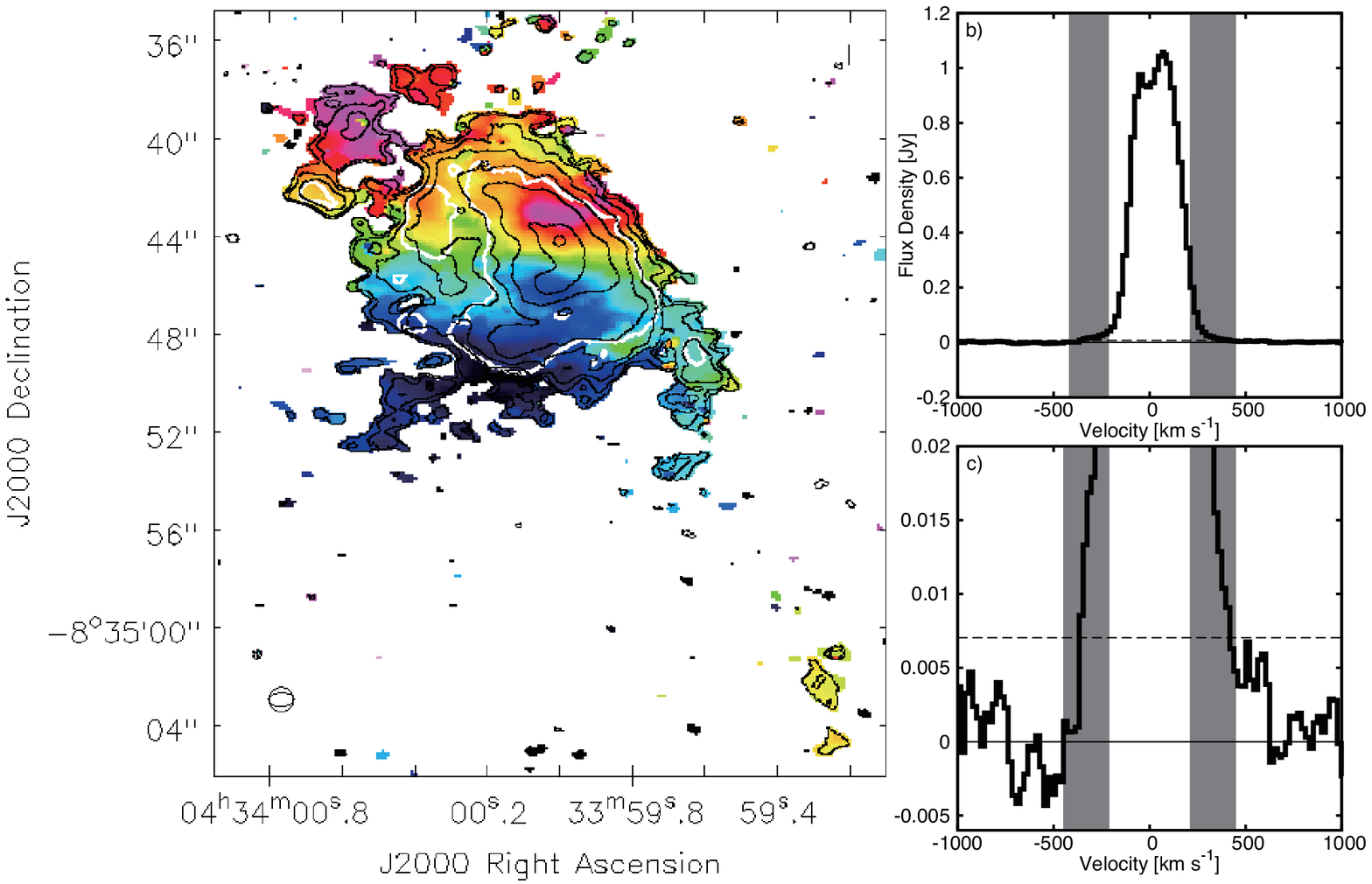}
\caption{(a) ACA+12~m-combined CO~(1--0) integrated intensity [black contour] and velocity field [color] images of NGC~1614.  The contours are 8.8 $\times$ (0.02, 0.04, 0.08, 0.16, 0.32, 0.64, and 0.96) Jy beam$^{-1}$ km s$^{-1}$. The velocity field in color scale ranges from 4500~km s$^{-1}$ to 4850~km s$^{-1}$.  The white contour shows the outline of the $uv$-clipped 12~m-only CO~(1--0) image at the level of 0.36 Jy beam$^{-1}$ km s$^{-1}$ (Figure~\ref{figure_2}a).  (b) CO~(1--0) spectrum toward the central 4\farcs6 aperture. (c) Zoomed-in CO~(1--0) spectrum of Figure~\ref{figure_2b}b.  The shaded area shows the CO line wings suggested by \citet{Garcia-burillo15}.
}
\label{figure_2b}
\end{center}
\end{figure*}

\section{Results} \label{results}
The $uv$-matched CO integrated intensity, velocity field, and velocity dispersion maps are shown in Figure~\ref{figure_2}, except for the ACA-combined CO~(1--0) image (Figure~\ref{figure_2b}a).
The channel maps are shown in Appendix~\ref{A1}.

\subsection{{\it uv}-matched CO Intensity, Velocity field, and Velocity Dispersion} \label{morphology}
The $uv$-matched CO distributions of NGC~1614 can be divided into two regions: the starburst ring (central 590~pc in radius) and an outer disk ($\sim$ 3.3~kpc in radius).
The starburst ring is detected in X-ray, polycyclic aromatic hydrocarbon \citep{Herrero-Illana14}, Pa $\alpha$ \citep{Alonso-Herrero01}, radio-to-FIR continuum (\citealt{Olsson10}; \citetalias{Saito15b}), HCN~(4--3), and HCO$^+$~(4--3) emission \citep{Imanishi&Nakanishi13, Sliwa14}.
Most of the CO emission in the starburst ring arise from the western side.
The asymmetry in the starburst ring coincides with the radio-to-FIR continuum and the HCO$^+$~(4--3) emission.
On the other hand, the outer disk is only detected in extended Pa $\alpha$ and soft X-ray emission \citep{Herrero-Illana14}, indicating a site of moderate star formation relative to the ring (but we still have missing flux for extended, low surface brightness CO~(6--5) emission).
All CO~(1--0) peaks outside the starburst ring coincide with the peaks detected in Pa $\alpha$ (Figure~\ref{figure_1}c).

\begin{figure*}
\begin{center}
\includegraphics[width=15cm]{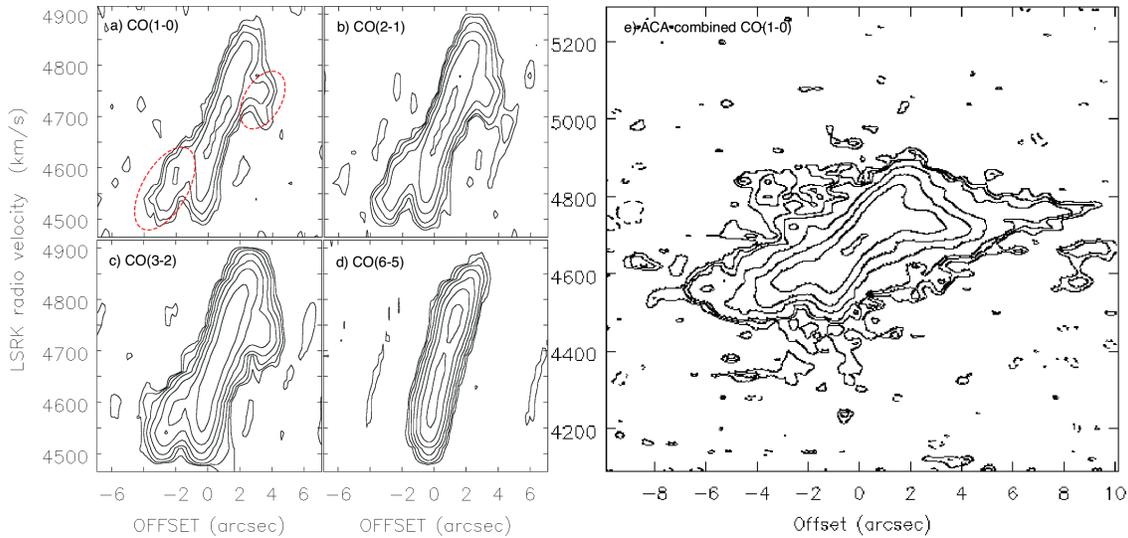}
\caption{(a/b/c/d) Spatially-averaged position-velocity diagram of each CO transition along the north-south direction (width = 13\farcs4) through the radio nucleus.  The $n$th contours are at 2$^{n}\sigma$ ($n$ = 1, 2, 3 ...).  $\sigma$ is the noise rms listed in Table~\ref{table_data}.  Red circles show the non-circular motion detected in lower-$J$ transitions. (e) Spatially-averaged position-velocity diagram of the ACA-combined CO~(1--0) along the north-south direction. The contours are 0.5 $\times$ (-2, 2, 3, 4, 8, 16, 32, and 64) mJy beam$^{-1}$.
}
\label{figure_5}
\end{center}
\end{figure*}

\begin{figure*}[tbh]
\begin{center}
\includegraphics[width=15cm]{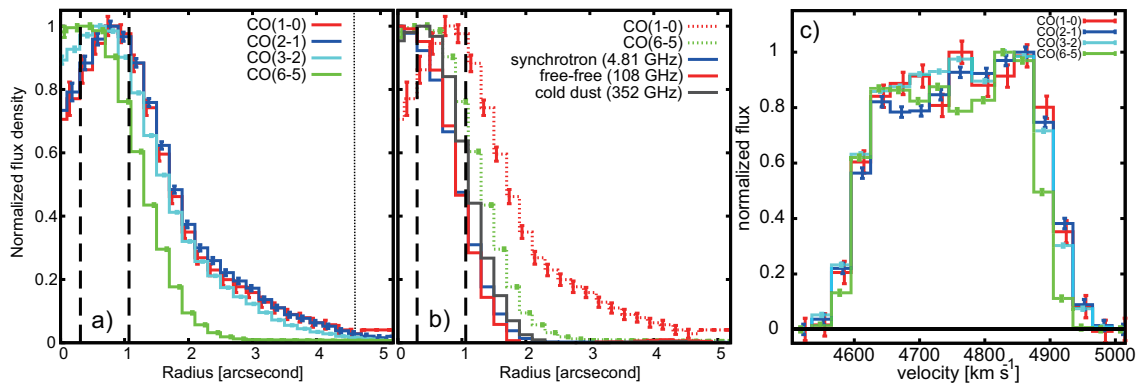}
\caption{(a) Azimuthally-averaged normalized radial distribution of each CO transition of NGC~1614 centered at the radio nucleus \citep{Olsson10}.  We only consider the statistical error.  The dashed lines show the approximate inner and outer radii of the nuclear starburst ring \citep{Xu15}.  The dotted line shows the approximate outer radius of the field of view of the CO~(6--5) data.  (b) Comparison between each CO transition (dotted line) and each continuum emission \citepalias[solid line;][]{Saito15b}.  All the data for the different CO transitions are convolved to the same resolution of 1\farcs0.  (c) Normalized CO Spectra of NGC~1614 toward the central 4\farcs6 aperture.
}
\label{figure_3}
\end{center}
\end{figure*}

The overall velocity structure of all the CO transitions seems to be dominated by rotation.
This is consistent with the velocity modeling with tilted concentric rings provided by \citet{Ueda14}.
The velocity field and dispersion of the starburst ring smoothly connect to those of the outer disk, which indicates that the kinematics (e.g., position angle, inclination) are similar.
We identify non-circular motions in a region with a length of 9\arcsec ($\sim$ 3~kpc) that is extended from north to south of the disk (dashed lines in Figure~\ref{figure_2}b) in all CO images, except for the CO~(6--5).
The apparent ``S"-shape feature in the velocity field suggests the presence of non-circular motions \citep[e.g.,][]{Kalnajs78} in the region where the dust lane connects with the starburst ring (Figure~\ref{figure_1}b) or a warped disk \citep[e.g., Centaurus A;][]{Espada12}.
The apparent characteristics are similar to the nearby barred spiral galaxy NGC~1097 which is a minor merger system \citep{Lin13}.

We investigate the position-velocity diagrams (PVDs) per the different CO transitions along the north-south direction with a width of 13\farcs4, length of 15\farcs0 through the nucleus (Figure~\ref{figure_5}a, \ref{figure_5}b, \ref{figure_5}c, and \ref{figure_5}d).
The width and length of these PVDs covers all of the CO emission (Figure~\ref{figure_2}).
The PVDs show rigid rotation (i.e., starburst ring) in all CO transitions and the ``S"-shape feature (dashed ellipses in Figure~\ref{figure_5}a) except for the CO~(6--5).
The direction of the PVD cut is similar to the kinematical major axis determined from the CO~(2--1) velocity field \citep[352\degr $\pm$ 1\degr;][]{Ueda14}.

While the large scale kinematics can be explained by both circular and non-circular motion, the central region appears to show outflows.
\citet{Garcia-burillo15} found high velocity CO~(1--0) wings (210 km s$^{-1}$ $<$ $|v - v_{\rm{sys}}|$ $<$ 450 km s$^{-1}$, where $v_{\rm{sys}}$ = 4763~km~s$^{-1}$) with the P.A. of 82$\degr$ in the central region.
They suggested these components are a putative bipolar outflow with an outflow rate of $\sim$ 40 $M_{\odot}$ yr$^{-1}$.
The direction and the outflow rate is similar to those of the ionized gas outflow \citep{Bellocchi12}.
We did not detect the line wings in all CO transitions, although they are detected in the ACA-combined CO~(1--0) image (see Section~\ref{ACA_result}).

In Figure~\ref{figure_3}a, we show the normalized radial distribution of the CO SLED of NGC~1614.
The CO~(6--5) emission is centrally concentrated, while the lower-$J$ CO transitions show an extended component (i.e., the outer disk).
The peak radius shifts toward outside as the CO transition shifts to lower-$J$, showing a radial excitation gradient in the starburst ring (see Section~\ref{RADEX}).
In Figure~\ref{figure_3}b, as a comparison, we plot the normalized radial distribution of the radio-to-FIR SED of NGC~1614 with the same angular resolution and MRS \citepalias{Saito15b}.
All of the continuum emission (synchrotron, free-free, and cold dust) are more compact than the lower-$J$ CO transitions, but the sizes are similar to the extent of the CO~(6--5) emission.
This suggests that the CO~(6--5) is a better tracer of star-forming activity than lower-$J$ CO transitions \citep[e.g.,][]{Xu15}.
Although differences are seen in the spatial distribution between the transitions, there are no significant differences in the velocity profile within the central 2\farcs3 radius as shown in Figure~\ref{figure_3}c, supporting that the rigid rotating component dominates the nuclear region.

\subsection{ACA+12~m-combined CO~(1--0) Data} \label{ACA_result}

We show the CO~(1--0) image that is generated by combining the ACA data with the 12~array data in Figure~\ref{figure_2b}a.
The global velocity field does not change significantly, although additional extended structures toward the north-east (redshifted) and south-west (blueshifted) appear.
These arm-like structures are located along the optical dust lanes, as already found by ACA-only CO~(1--0) image \citep{Konig16}, and Pa $\alpha$ (Figure~\ref{figure_1}c).
The low S/N high-velocity wings (Figure~\ref{figure_2b}b and \ref{figure_2b}c) are consistent with the presence of a putative outflow suggested by \citet{Garcia-burillo15}.

\begin{figure*}
\begin{center}
\includegraphics[width=15cm]{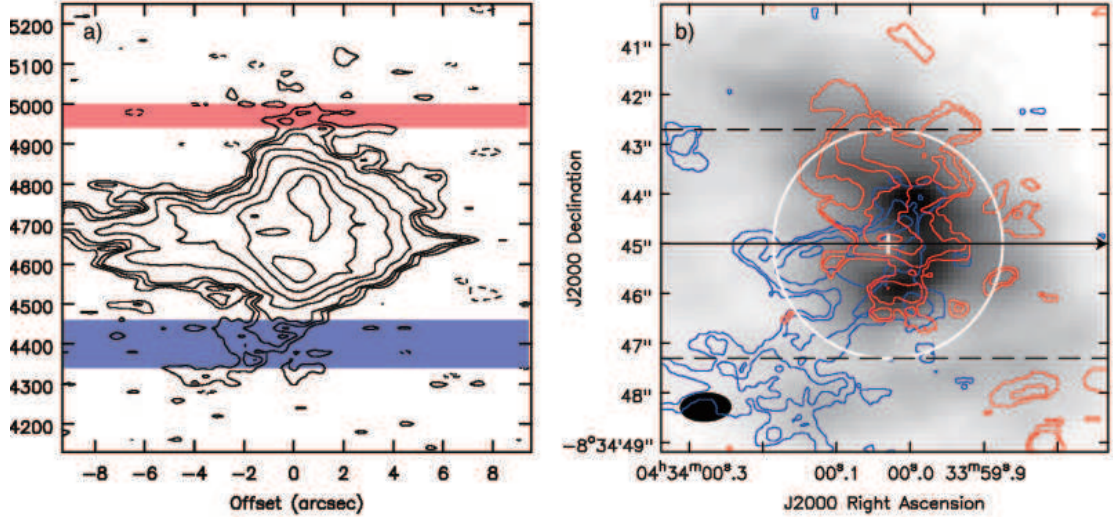}
\caption{(a) Spatially-averaged position-velocity diagram of the ACA-combined CO~(1--0) data along the east-west direction (width = 4\farcs6).  The contour are 0.5 $\times$ (-2, 2, 3, 4, 8, 16, 32, and 64) mJy beam$^{-1}$.  This PVD only contains velocity components between the dashed lines in Figure~\ref{figure_12b}b.  (b) Integrated intensity contour images of the high-velocity components \citep[i.e., molecular outflow suggested by][]{Garcia-burillo15} overlaid on the ACA-combined CO~(1--0) image (Figure~\ref{figure_2b}a) with grey scale.  The blue and red contours show blueshifted and redshifted components, respectively.  The blue contours are 0.45 $\times$ (0.16, 0.32,0.64, and 0.96) Jy beam$^{-1}$ km s$^{-1}$, whereas the red contours are 0.32 $\times$ (0.16, 0.32,0.64, and 0.96) Jy beam$^{-1}$ km s$^{-1}$.  The white circle indicates the approximate outer edge of the starburst ring (\citealt{Konig13,Herrero-Illana14, Xu15}; \citetalias{Saito15b}).
}
\label{figure_12b}
\end{center}
\end{figure*}

We investigate the PVD of the ACA-combined image along the north-south (along the kinematical major axis with a width of 13\farcs4; Figure~\ref{figure_5}e) and east-west direction (along the kinematical minor axis with a width of 4\farcs6; Figure~\ref{figure_12b}a) through the radio nucleus \citep{Olsson10}.
Along the major axis, we found more extended components than the $uv$-clipped CO~(1--0) image.
The ACA data shows features that are consistent with bar-induced inflow motions predicted by \citet{Iono04a}.
In contrast, as shown in the PVD along the minor axis which emphasizes low surface brightness components within the 4\farcs6 width, there are high-velocity components around the nuclear region.
Since the high-velocity components show the same spatial and spectral distributions as the molecular gas outflow suggested by \citet{Garcia-burillo15}, we regard the high-velocity components as the molecular gas outflow.
We constructed the integrated intensity image of the blueshifted (4340 - 4460~km s$^{-1}$) and redshifted (4940 - 5000~km s$^{-1}$) outflows in Figure~\ref{figure_12b}b.
The extracted velocity ranges contain high-velocity wings stronger than 3$\sigma$.
The CO~(1--0) outflow shows an extended blueshifted (redshifted) emission toward the south-east (north-west) direction.
The molecular gas outflow seems to encompass the ionized gas outflow \citep{Bellocchi12}.
This picture is consistent with the molecular gas outflow entrained by the ionized gas outflow in the nearby starburst galaxy NGC~253 and M51 \citep{Bolatto13N,Querejeta16}.

\begin{figure*}
\begin{center}
\includegraphics[width=18cm]{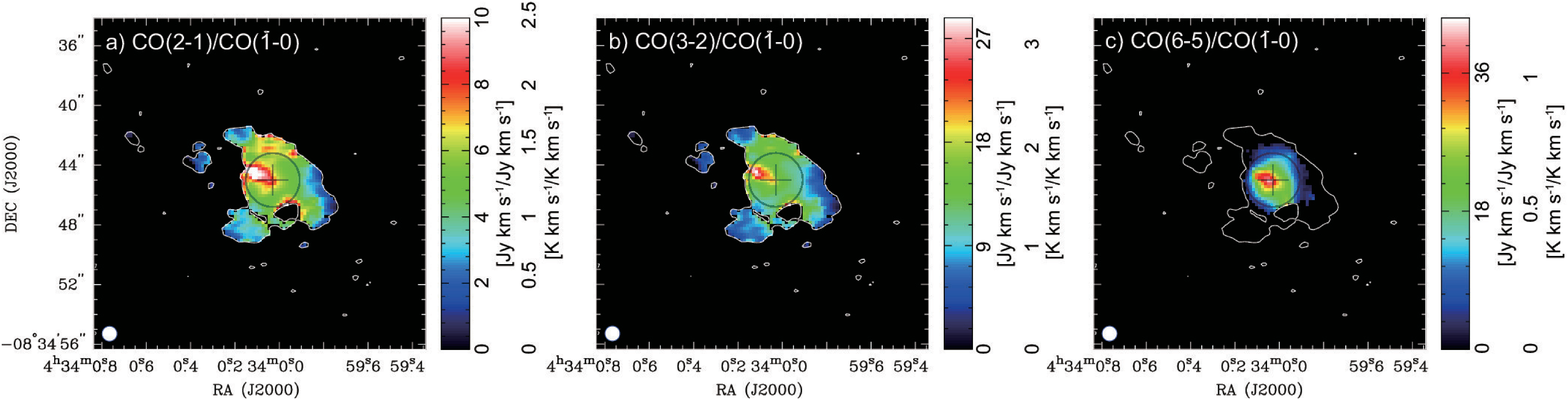}
\caption{(a) Image of the $uv$-matched CO~(2--1)/CO~(1--0) intensity ratio for NGC~1614.  The ratio in color scale ranges from 0 to 10 in Jy km s$^{-1}$/(Jy km s$^{-1}$)$^{-1}$ (0 to 2.5 in K km s$^{-1}$/(K km s$^{-1}$)$^{-1}$).  The contour shows the CO~(1--0) outline (0.54 Jy beam$^{-1}$ km s$^{-1}$).  The black cross indicates the nucleus which is detected in Pa $\alpha$ and the radio continuum emission \citep{Herrero-Illana14}.  The black circle indicates the approximate outer edge of the starburst ring (\citealt{Konig13,Herrero-Illana14, Xu15}; \citetalias{Saito15b}).  (b)  Image of the $uv$-matched CO~(3--2)/CO~(1--0) intensity ratio.  The ratio in color scale ranges from 0 to 28.8 in Jy km s$^{-1}$/(Jy km s$^{-1}$)$^{-1}$.  (c)  Image of the $uv$-matched CO~(6--5)/CO~(1--0) intensity ratio.  The ratio in color scale ranges from 0 to 43.2 in Jy km s$^{-1}$/(Jy km s$^{-1}$)$^{-1}$.
}
\label{figure_4}
\end{center}
\end{figure*}

\begin{figure*}
\begin{center}
\includegraphics[width=15cm]{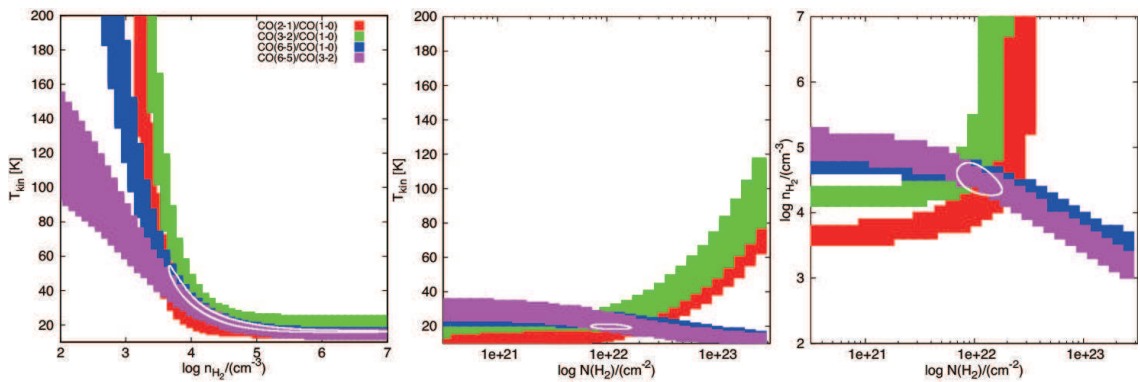}
\caption{Results of the single-phase RADEX modeling at the outer region contained within a radius of 2\farcs8 - 3\farcs0 (924~pc - 990~pc): (a) log $n_{\rm H_2}$ - $T_{\rm kin}$ plane, (b) log $N$(H$_2$) - $T_{\rm kin}$ plane, and (c) log $N$(H$_2$) - log $n_{\rm H_2}$ plane.  All planes are cut through the best-fit point.  Each colorized area corresponds to the 1$\sigma$ range of each observed line ratio.  The white outline shows the 90\% confidence contour of the fit.
}
\label{figure_14}
\end{center}
\end{figure*}

\subsection{CO Line Ratios} \label{ratio}
We show CO~(2--1)/CO~(1--0), CO~(3--2)/CO~(1--0), and CO~(6--5)/CO~(1--0) line ratio images in Figure~\ref{figure_4}a, b, and c, respectively.
We divide the integrated intensity maps (Figure~\ref{figure_2}d, g, and j) by the CO(1-0) integrated intensity map (Figure~\ref{figure_2}a) to make the line ratio maps.
After that, we masked all the line ratio maps by using the CO~(1--0) outline which is shown in Figure~\ref{figure_4}.
The CO~(2--1)/CO~(1--0) and CO~(3--2)/CO~(1--0) line ratios show higher values than the optically-thick thermalized value, while the CO~(6--5)/CO~(1--0) line ratio is moderate.
Those high CO~(2--1)/CO~(1--0) and CO~(3--2)/CO~(1--0) line ratios indicate low optical depths for the CO~(1--0) line \citep[e.g.,][]{Downes&Solomon98} and/or the presence of multiple phase ISM \citep[e.g., CO~(3--2)/CO~(1--0) ratio in Arp~220;][]{Rangwala11}.
The peak position of each line ratio is located at 1\arcsec east of the nucleus.
Since it coincides with the eastern part of a possible ionized and molecular gas outflow \citep{Bellocchi12, Garcia-burillo15}, the high line ratio component might be due to the interaction between the outflow and the starburst ring.
Strong outflow can heat surrounding ISM kinematically.
This results in highly excited (i.e., low-$J$ depopulation) or highly turbulent (i.e., low optical depth) molecular gas around the shock interface, and thus high line ratios.
High CO~(3--2)/CO~(1--0) ratio is also observed in the circumnuclear disk (CND) of Seyfert 2 galaxy NGC~1068 \citep[See Section~7.2 of][]{Garcia-burillo14}.
The high ratio in the CND are located parallel to the bipolar jet, which may indicate that the CND is affected by the nuclear jet.
A similar situation is found in the nuclear region of NGC~1614 (Figure~\ref{figure_4}), which may indicate outflow-ring interaction.
However, further molecular gas and ionized gas observations are necessary to confirm this scenario quantitatively.

\section{CO SLED Modeling} \label{model}
The CO SLEDs in U/LIRGs are often explained by a combination of CRDR, PDR, XDR, and MDR.
However, especially for NGC~1614, previous studies suggested that contributions from CRDR and XDR are small or those models cannot reproduce observed quantities \citep[see][in details]{Herrero-Illana14,Xu14}, so we only focus on the effect of both PDR and MDR in this Paper.

\subsection{Single-phase RADEX Model} \label{RADEX}

We used the non-LTE radiative transfer code RADEX \citep{van_der_Tak07} and varied the parameters until the residuals between the observed line fluxes and the modeled line fluxes are minimized in a $\chi^2$ sense.
Assuming a single-phase ISM (i.e., the gas physics can be represented by a single set of the excitation parameters), an expanding sphere geometry ($dv$ = 200 km s$^{-1}$), the cosmic microwave background temperature ($T_{\rm{bg}}$ = $T_{\rm CMB}$ = 2.73~K), and a [CO]/[H$_2$] abundance (3 $\times$ 10$^{-4}$) which is similar to the standard value observed in Galactic molecular clouds \citep{Blake87}, we derived the physical conditions of molecular gas.
The upper state energies and the Einstein coefficients were taken from the {\it Leiden Atomic and Molecular Database} \citep[LAMDA;][]{Schoier05}.
We varied the gas kinetic temperature within a range of $T_{\rm{kin}}$ = 10 - 300~K (d$T_{\rm{kin}}$ = 1~K), the gas density of $n_{\rm{H_2}}$ = 10$^2$ - 10$^7$~cm$^{-3}$ (d$n_{\rm{H_2}}$ = 10$^{0.1}$~cm$^{-3}$), and the gas column density of $N$(H$_2$) = 10$^{20.5}$ - 10$^{23.5}$~cm$^{-2}$ (d$N$(H$_2$) = 10$^{0.1}$~cm$^{-2}$).
For the input parameters, we used the radial flux distribution of each CO line shown in Figure~\ref{figure_3}, which were radially binned at 0\farcs2 intervals.

We note that the $T_{\rm bg}$ = $T_{\rm CMB}$ is a reasonable assumption although one might expect that, given the intense dusty star formation, a higher $T_{\rm bg}$ is more suitable (i.e., $T_{\rm bg}$ $\sim$ $T_{\rm dust}$).
We estimated the dust opacity at the starburst ring in \citet{Saito15b} and found a low dust opacity ($<$ 0.21) even at 690 GHz.
Therefore, at this low frequency regime (115-690 GHz), the background temperature cannot be as high as $T_{\rm dust}$.

\begin{figure}
\begin{center}
\includegraphics[width=8.5cm]{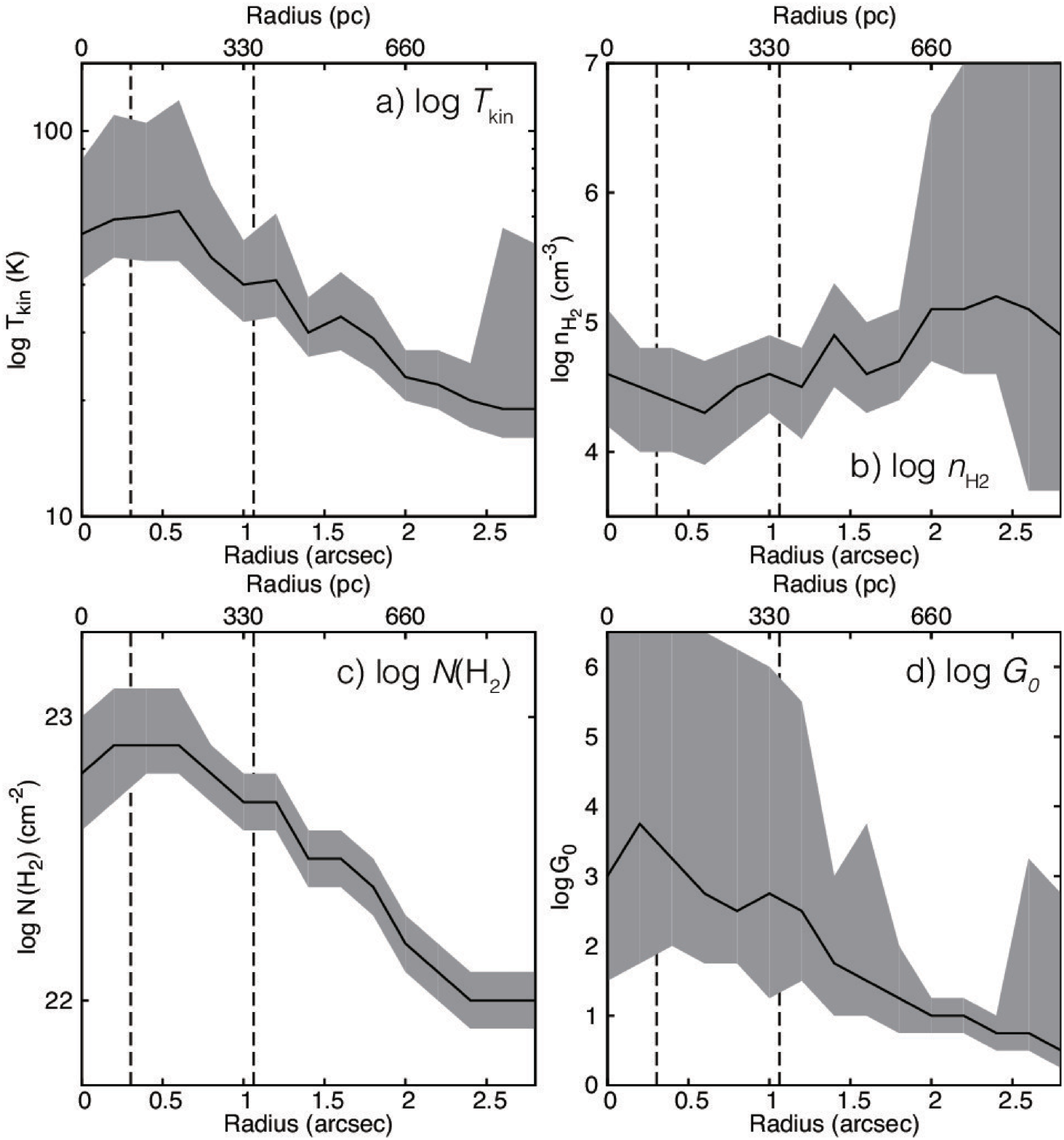}
\caption{Results of the RADEX and PDR Toolbox modeling as a function of the radius;  (a) kinetic temperature, (b) molecular gas density, (c) column density of molecular hydrogen, and (d) far-UV radiation field.  The dashed lines show the approximate inner and outer radii of the nuclear starburst ring \citep{Xu15}.  The shaded areas show the 90\% confidence.
}
\label{figure_6}
\end{center}
\end{figure}

\begin{figure}
\begin{center}
\includegraphics[width=6cm]{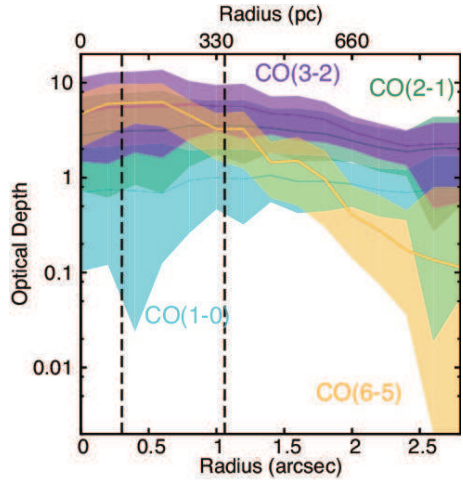}
\caption{Optical depth of each CO transition as a function of the radius.  The dashed lines show the approximate inner and outer radii of the nuclear starburst ring \citep{Xu15}.  The shaded areas show the 90\% confidence.
}
\label{figure_7}
\end{center}
\end{figure}

We show results of the RADEX modeling at the outer region contained within a radius of 2\farcs8 - 3\farcs0 in Figure~\ref{figure_14}.
Line ratios related to CO~(6--5) show different distributions to lower-$J$ CO line ratios in the parameter planes, demonstrating the importance of the CO~(6--5) data.
The results as a function of radius are shown in Figure~\ref{figure_6}a, \ref{figure_6}b, and \ref{figure_6}c.
The minimum reduced-$\chi^2$ of the central 1\farcs8 is relatively poor ($\gtrsim$ 3.6), whereas that of the outer region (r $>$ 1\farcs8) is $\lesssim$ 3.6.
Gas density is high ($\sim$ 10$^{5}$~cm$^{-3}$) independently of the distance from the nucleus.
On the other hand, both the $T_{\rm{kin}}$ (= 20 - 70~K) and $N$(H$_2$) (= 10$^{22.0 - 22.9}$~cm$^{-3}$) decrease as the radius increases.
Their peaks coincide with the radius of the starburst ring \citep[100 - 350~pc;][]{Konig13,Xu15}.
The derived $T_{\rm{kin}}$ are about two times higher than the results of a similar modeling for NGC~1614 by \citet{Sliwa14}.
This can be explained by the different spatial sampling between our analysis and theirs.
In order to match the MRS between data sets, \citet{Sliwa14} used single-dish data to compensate for the missing short baselines of their interferometric CO~(2--1) and CO~(3--2) data, while we clipped the inner $uv$ data at 45~k$\lambda$ (see Section~\ref{obs}).

We calculated the optical depth for each transition (Figure~\ref{figure_7}), yielding that $\tau_{\rm{CO(1-0)}}$ shows moderate optical depth ($\lesssim$ 1), $\tau_{\rm{CO(2-1)}}$ and $\tau_{\rm{CO(3-2)}}$ are optically thick (1 - 10), and $\tau_{\rm CO(6-5)}$ becomes thin as the radius increases ($\tau_{\rm CO(6-5)}$ = 8 to 0.1).
The derived $\tau_{\rm{CO(1-0)}}$ for NGC~1614 is consistent with that is estimated for a mid-stage merger VV~114 \citep[$\sim$ 1;][]{Saito15}, the central region of a close galaxy pair NGC~6240 \citep[0.2 - 2;][]{Iono07}, the overlap region of the mid-stage merger Arp~299 \citep[0.5 - 1.5;][]{Sliwa12}, and the local Mid-stage Merger the Antennae \citep[0.2 - 1.5;][]{Zhu03}.
This suggests that luminous mergers have CO~(1--0) line of moderate optical depths and thick high-$J$ CO condition \citep[e.g.,][]{Downes&Solomon98}.

\begin{figure*}
\begin{center}
\includegraphics[width=18cm]{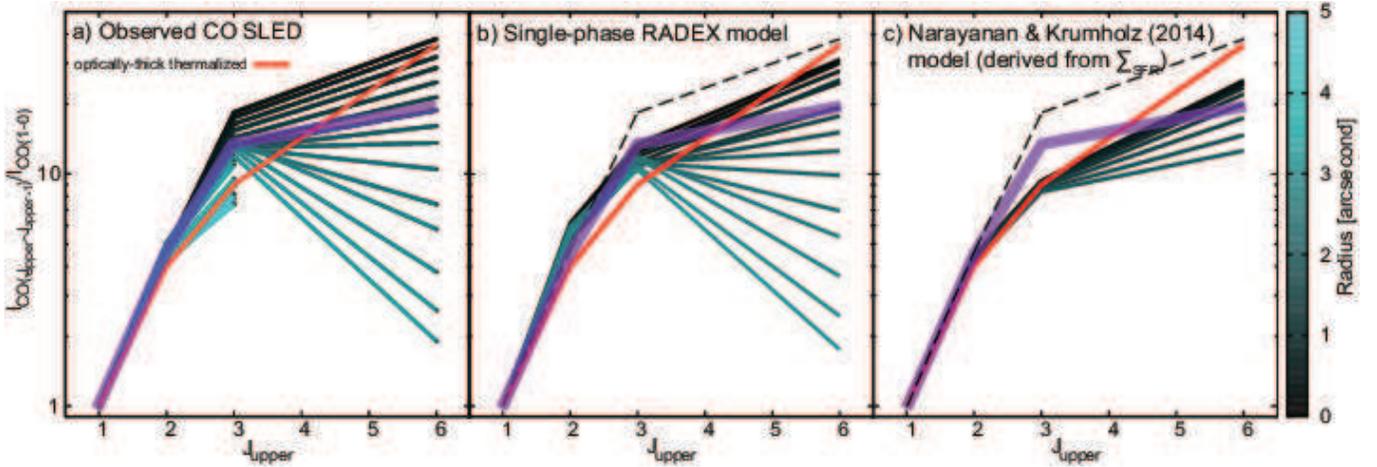}
\caption{(a) Observed CO spectral line energy distributions (SLEDs) of NGC~1614 for concentric rings with the radius from 0\farcs2 to 5\farcs0.  The color scale ranged from 0\farcs0 to 5\farcs0 (0\farcs2 step from top to bottom lines) shows the radius of the concentric rings which are used for the photometry.  The red line indicates a thermalized and optically-thick CO SLED as a reference.  CO SLED averaged over inside 1\farcs8 radius is shown in the purple line.  (b) Modeled CO SLED reconstructed by the single-phase RADEX. The dashed line shows the observed CO SLED within the central 0\farcs2 radius.  (c) Modeled CO SLED reconstructed from the \citet{Narayanan&Krumholz14} model.  The $\Sigma_{\rm{SFR}}$ derived from the radio-to-FIR SED \citepalias{Saito15b} is used as the input parameter.
}
\label{figure_8}
\end{center}
\end{figure*}

We reconstructed the best-fit CO SLED from models in order to compare with the observational data (Figure~\ref{figure_8}a and \ref{figure_8}b).
The CO SLEDs in the outer disk ($>$ 1\farcs8) show good agreement with the observations, while those in the starburst ring ($\lesssim$ 1\farcs8) systematically underestimated the CO~(3--2)/CO~(1--0) and CO~(6--5)/CO~(1--0) ratios.
One possibility to explain the systematic difference is that the starburst ring of NGC~1614 has an additional warmer ISM, as suggested by \citet{Sliwa14}.
The additional warm gas can move the modeled CO~(3--2)/CO~(1--0) and CO~(6--5)/CO~(1--0) ratios toward a higher value by fixing the lower-$J$ CO fluxes.

We note that the flux differences between the observations and the model are unlikely to be caused by dust extinction.
Although some U/LIRGs show high dust opacities even in the submillimeter wavelengths \citep[e.g.,][]{Rangwala11, Wilson14}, the cold dust in the starburst ring of NGC~1614 is still optically thin at 691~GHz \citepalias[0.06 - 0.21;][]{Saito15b}.

\subsection{Comparing with the Narayanan \& Krumholz Model} \label{nk}

We compare the observed CO SLED (Figure~\ref{figure_8}a) with the model of \citet{Narayanan&Krumholz14}, which can parameterize the CO SLED of galaxies by a power-law function of $\Sigma_{\rm{SFR}}$.
Using the $\Sigma_{\rm{SFR}}$ derived by \citetalias{Saito15b}, we reconstructed the CO SLED (Figure~\ref{figure_8}c).
The $\Sigma_{\rm{SFR}}$ were derived by fitting the radial radio-to-FIR SEDs binned at 0\farcs2 intervals up to 1\farcs8 radius assuming the absence of a strong or a hearvily obscured AGN.
Each continuum images have the same angular resolution of 1\farcs0 and MRS of 4\farcs6 as the CO images, so we can compare the observed CO SLEDs with the reconstructed CO SLEDs dicrectly.

As clearly seen in Figure~\ref{figure_8}a and \ref{figure_8}c, this model is inconsistent with the observed CO SLEDs at the starburst ring ($<$ 1\farcs8).
The CO~(3--2)/CO~(1--0) and CO~(6--5)/CO~(1--0) ratios are also underestimated as already described in Section~\ref{RADEX}.
A multi-phase ISM model is one of the possible explanations.
Because our analysis includes only four CO transitions and can only constrain a small number of free parameters, we will focus on a two-phase ISM model rather than a multiple-phase ISM (Section~\ref{ISM}).
The $\Sigma_{\rm{SFR}}$ used here is extinction-free because \citetalias{Saito15b} used the low frequency continuum from 4.81 to 691~GHz.
Thus, we can ignore the extinction effect in this comparison.

\subsection{PDR Model} \label{PDR}
We used the PDR Toolbox \citep{Kaufman06}, which can model the beam-averaged incident far-ultraviolet (FUV; 6~eV $<$ $E$ $<$ 13.6 eV) radiation field intensity ($G_{\rm{0}}$ in the unit of 1.6 $\times$ 10$^{-3}$ erg cm$^{-2}$ s$^{-1}$) and PDR hydrogen nucleus density ($n$ cm$^{-3}$ = $n_{\rm{H}}$ + 2$n_{\rm{H_2}}$) from the flux density of each CO transition, in order to better understand the starburst activities of NGC~1614.
We varied $n$ within the 90\% confidence range of 2$n_{\rm{H_2}}$ derived by the single-phase RADEX (d$n$ = 10$^{0.25}$~cm$^{-3}$) and $G_{\rm{0}}$ within a range of 10$^{-0.50}$ - 10$^{6.50}$ (d$G_{\rm{0}}$ = 10$^{0.25}$).
Here we assumed that the $n_{\rm{H}}$ is negligible compared to the 2$n_{\rm{H_2}}$ term because the central kpc of mergers are often dominated by the molecular phase \citep[e.g.,][]{Iono05, Kaneko05}.
This assumption is consistent with the compact CO distributions (Figure~\ref{figure_2}) and the extended arc-like HI distribution (Figure~\ref{figure_1}a) around the kpc region of NGC~1614.
The results are shown in Figure~\ref{figure_6}d.
The minimum reduced-$\chi^2$ is $\sim$ 1 for both grids.
The starburst ring has higher $G_{\rm{0}}$ ($>$ 10$^{1.5}$, average = 10$^{2.7}$) than the outer disk ($\lesssim$ 10$^{1.5}$, average = 10$^{0.9}$).
The average value of the whole galaxy is $\sim$ 10$^{2.3}$.

In order to check the reliability of the derived $G_{\rm{0}}$, we roughly estimate the averaged FUV radiation field on dust-obscured star-forming region using the prescription provided by \citet{Papadopoulos14},

\begin{eqnarray}
G_{\rm{0}}(L_{\rm{IR}})\sim&&3\times10^2\lambda_{*}({\rm pc})\left(\frac{L_{\rm{IR}}}{10^{10}\:L_{\odot}}\right)\nonumber\\
&&\times\left(\frac{R_{\rm{SB}}}{100\:{\rm pc}}\right)^{-3}[1 - e^{-(R_{\rm{SB}}/\lambda_{*})}],
\end{eqnarray}
where $\lambda_{*}$ is the mean distance of FUV photons propagation before being absorbed, $R_{\rm{SB}}$ is the size of the starburst ring \citep[$\sim$ 350~pc; ][]{Xu15}.
For the clumpy PDR model, $\lambda_{*}$ is $\lesssim$ 1~pc \citep{Meixner&Tielens93}.
The derived $G_{\rm{0}}(L_{\rm{IR}})$ is $\lesssim$ 10$^{2.5}$, which is consistent with the average $G_{\rm{0}}$ derived from the CO SLED.

\section{Discussion} \label{discuss}
\subsection{The ``S''-shape velocity field} \label{inflow}
The ``S"-shape morphology in the velocity field suggests the presence of a north-south bar or a warped disk morphology.
It coincides with the leading edge of the main near-IR bar suggested by \citet{Olsson10}, so we suggest that the north-south bar results in the ``S"-shape non-circular motions (i.e., inflow along a bar) detected in the lower-$J$ CO transitions.
The presence of the cold gas inflow along the bar is also suggested by \citet{Olsson10}.
In addition, analysis of N-body/SPH simulation performed by \citet{Iono04a} revealed that tidal interaction between gas-rich disks can produce gas flows directly toward the central regions due to the production of both transient arms and long-lived $m$ = 2 bars (i.e., bar instability).
They also found that the inflowing gas is characterized by distinct diffuse gas clumps in the PVDs, while dense gas only present in the central rigid rotating component.
This simulation result is consistent with our observations including non-detection of the ``S"-shape feature in the CO~(6--5) image.

\begin{figure}
\begin{center}
\includegraphics[width=8.5cm]{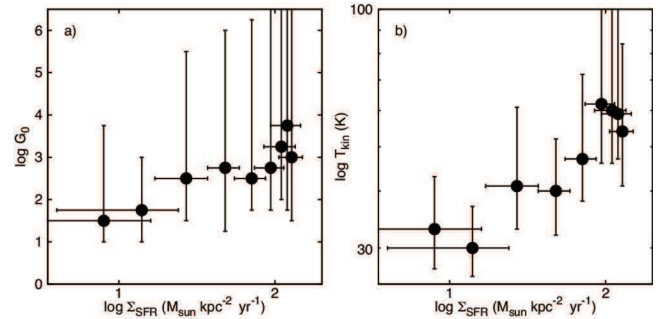}
\caption{(a) Plot of the star formation rate surface density vs. the FUV radiation field at each concentric ring.  The errors correspond to the 90\% confidence.  (b) Plot of the star formation rate surface density vs. the kinetic temperature at each concentric ring.
}
\label{figure_9}
\end{center}
\end{figure}

\subsection{Power Source of the Single-phase ISM} \label{power}

We compare $T_{\rm{kin}}$ and $G_0$ with $\Sigma_{\rm{SFR}}$, which is derived by modeling the radio-to-FIR SED \citepalias{Saito15b}, in Figure~\ref{figure_9}.
There is a positive correlation with a correlation coefficient of $r$ = 0.91 between $\log \Sigma_{\rm{SFR}}$ and $\log G_0$.
This is a sensible result because both parameters are related to star formation.
The relation between $\log \Sigma_{\rm{SFR}}$ and $\log T_{\rm{kin}}$ also shows a positive correlation ($r$ = 0.93).
Using a PDR model described in equation 3 of \citet{Papadopoulos14} and assuming $T_{\rm{kin}} \gtrsim T_{\rm{dust}}$ which is found by PDR models \citep[e.g., ][]{Hollenbach&Tielens99}, the derived $T_{\rm{kin}}$ of 20 - 70 K corresponds to $\log G_0$ of $\lesssim$ 10$^{1.4 - 4.6}$.
This is consistent with the result of the CO SLED model described in Section~\ref{PDR}.
These models and a simple comparison indicates that the dusty starburst in NGC~1614 is enough to power the CO excitation up to $J_{\rm{upp}}$ = 6 even if we just assume a single-phase molecular gas ISM for the starburst ring.

\subsection{Spatially resolved CO-to-H$_2$ Conversion Factor} \label{conversion}

Molecular hydrogen mass is one of the most fundamental parameters to characterize galaxies in relation to their starbursts and AGNs.
Although the galactic-scale CO~(1--0) luminosity to H$_2$ mass conversion factor ($M_{\rm{H_2}}$/$L'_{\rm CO(1-0)}$ $\equiv$ $\alpha_{\rm{CO(1-0)}}$) in nearby U/LIRGs has been well studied \citep[][for a review]{Bolatto13}, the spatial properties are not fully understood.

We measure the $\alpha_{\rm{CO(1-0)}}$ as a function of radius using,

\begin{figure}
\begin{center}
\includegraphics[width=6cm]{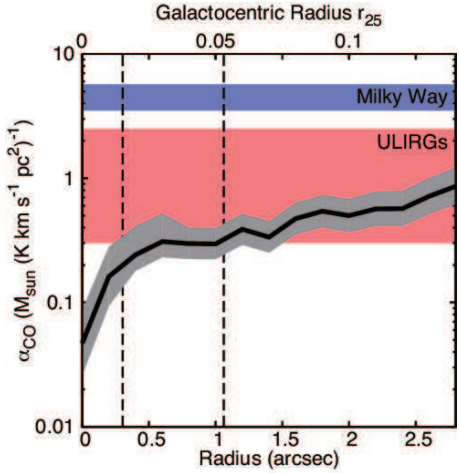}
\caption{The CO-to-H$_2$ conversion factor (= $\alpha_{\rm{CO(1-0)}}$ = $M_{\rm{H_2}}$/$L'_{\rm{CO(1-0)}}$) as a function of the radius derived by the signle-phase RADEX model (i.e., $N$(H$_2$)).  The dashed lines show the approximate inner and outer radii of the nuclear starburst ring \citep{Xu15}.  The shaded areas show the 90\% confidence.  The blue and red regions show approximate $\alpha_{\rm{CO(1-0)}}$ for the Milky Way and ULIRGs \citep{Bolatto13}.
}
\label{figure_10}
\end{center}
\end{figure}

\begin{figure}
\begin{center}
\includegraphics[width=6cm]{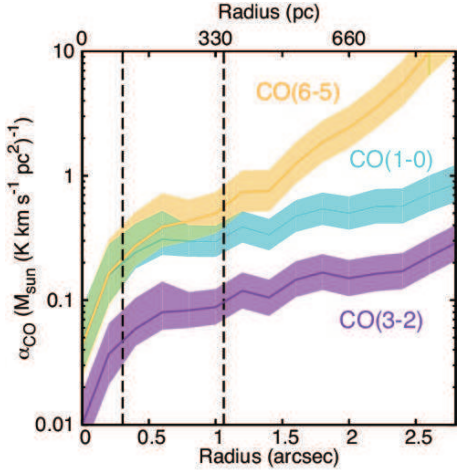}
\caption{The CO-to-H$_2$ conversion factor for CO~(1-0), CO~(3--2), and CO~(6--5) lines as a function of the radius.  The dashed lines show the approximate inner and outer radii of the nuclear starburst ring \citep{Xu15}.  The shaded areas show the 90\% confidence.
}
\label{figure_10b}
\end{center}
\end{figure}

\begin{equation}
\alpha_{\rm{CO(1-0)}} = 1.36\frac{m_{\rm H_2}\:N({\rm H_2})\:S_{\rm{ring}}}{L'_{\rm CO(1-0)}}\:M_{\odot}\:({\rm K\:km\:s^{-1}\:pc^2})^{-1},
\end{equation}
where $m_{\rm H_2}$ is the mass of the hydrogen molecule, $N({\rm H_2})$ is the column density derived by RADEX (Section~\ref{RADEX}), $S_{\rm{ring}}$ is the area of the concentric rings, and $L'_{\rm CO(1-0)}$ is the CO~(1--0) luminosity.
We multiply 1.36 to account for the helium abundance relative to hydrogen.
The derived $\alpha_{\rm{CO(1-0)}}$ at the starburst ring is 0.2 - 0.5 ($\frac{3 \times 10^{-4}}{\rm{[CO]/[H_2]}}$) except for the central hole (central 0\farcs2 radius) which yields $\sim$ 0.05 ($\frac{3 \times 10^{-4}}{\rm{[CO]/[H_2]}}$) (Figure~\ref{figure_10}).
This is lower than 0.9 - 1.5 ($\frac{3 \times 10^{-4}}{\rm{[CO]/[H_2]}}$) at the starburst ring and 0.1 ($\frac{3 \times 10^{-4}}{\rm{[CO]/[H_2]}}$) at the central hole derived by \citet{Sliwa14}.
The difference between both models can be explained by the presence of virialized giant molecular clouds (GMCs) and the different treatment of the missing flux.
Assuming that GMCs in the central region of NGC~1614 are virialized, $\alpha_{\rm{CO(1-0)}}$ depends on gas density and kinetic temperature \citep[$\alpha_{\rm{CO(1-0)}} \propto \rho^{0.5} T_{\rm{kin}}^{-1}$ where $\rho$ is the gas density;][]{Solomon&Vanden_Bout05, Bolatto13}.
As a consequence of the different missing flux treatment, the results of our RADEX modelings for the starburst ring are warmer (27 - 120~K and 10$^{3.9 - 5.3}$ cm$^{-3}$) than the results in \citet{Sliwa14} (19 - 64~K and 10$^{3.0 - 6.5}$ cm$^{-3}$) (see also Section~\ref{RADEX}).
Assuming the same gas density ($\rho$) and beam filling factor for both models, $\alpha_{\rm{CO(1-0)}}$ derived from our model shall indeed be $\sim$ 2 times lower than the \citet{Sliwa14} value, indicating that both models are consistent.

We note that the treatment of the missing flux (i.e., how to achieve the same MRS for all transitions) should be different depending on the situation.
When we have short-spacing data for ``all" transitions, we should combine the data with data obtained with the compact configurations, which allows us to recover extended structures.
In the case of this Paper, single-dish or ACA data for several transitions are lacking, and thus we chose to clip the inner $uv$, making $\alpha_{\rm{CO(1-0)}}$ different from \citet{Sliwa14}.

Comparing to local LIRGs, it is known that nearby spirals show larger $\alpha_{\rm{CO(1-0)}}$ \citep[e.g.,][]{Bolatto13}.
\citet{Sandstrom13} constructed $\alpha_{\rm{CO(1-0)}}$ as a function of galactocentric radius ($r_{25}$ = $r$/$R_{25}$, where $R_{25}$ is the B-band isophotal radii at 25 mag arcsec$^{-2}$) for 26 spiral galaxies, and found lower $\alpha_{\rm{CO(1-0)}}$ at the galaxy centers ($r_{25}$ $<$ 0.2) by a factor of 1 $\sim$ 5 relative to their disks ($r_{25}$ = 0.4 - 1.0).
Using $R_{25}$ of 6.5~kpc ($\sim$ 20\arcsec) derived by $R_{25}$ = 1.9 $\times$ $R_{\rm{eff}}$ \citep{Bellocchi13} and $R_{\rm{}eff}$ = 3.44~kpc \citep{Ueda14} for NGC~1614, the outer radius of the central hole, the starburst ring, and the CO disk of NGC~1614 are $r_{25}$ $\sim$ 0.01, 0.05, and 0.14, respectively (Figure~\ref{figure_10}).
This indicates that NGC~1614 shows a similar radial $\alpha_{\rm{CO(1-0)}}$ trend to nearby spirals, but it has a more compact distribution and lower $\alpha_{\rm{CO(1-0)}}$ value than that of nearby spirals for a given galactocentric radius.

We calculate the conversion factor for CO~(3--2) ($M_{\rm{H_2}}$/$L'_{\rm CO(3-2)}$ $\equiv$ $\alpha_{\rm CO(3-2)}$) and CO~(6--5) ($M_{\rm{H_2}}$/$L'_{\rm CO(6-5)}$ $\equiv$ $\alpha_{\rm CO(6-5)}$) in Figure~\ref{figure_10b} assuming the single-phase ISM.
The radial distribution of $\alpha_{\rm CO(3-2)}$ is almost the same as that of $\alpha_{\rm CO(1-0)}$, while the $\alpha_{\rm CO(6-5)}$ shows a steeper shape, except for the nuclear region ($<$ 1\farcs0), which shows a similar value of $\alpha_{\rm{CO(1-0)}}$ because of the sub-thermalized conditions in the outer region.

\subsection{Two-phase Modeling for the Starburst Ring} \label{ISM}
\subsubsection{Two-phase ISM model} \label{two}
{\it Herschel} and other single-dish observations revealed that the CO SLED of nearby U/LIRGs can be represented by multi-phase molecular gas ISM model \citep[e.g.,][]{van_der_Werf10, Rangwala11}.
Although the spatial properties of the CO SLED for U/LIRGs are unclear because of the limited number of high-resolution CO data, our high-resolution CO~(1--0) and CO~(2--1) observations, as well as the archival CO~(3--2) and CO~(6--5) data, allow us to study this for a LIRG for the first time.

Before discussing the two-phase molecular gas ISM in NGC~1614, we note that the dust ISM in the starburst ring can be described by warm and cold components.
\citet{Pereira-Santaella15} reported the presence of a warm dust component in the starburst ring ($\sim$ 110~K) revealed by modeling of the mid-IR emission, although there is also a cold dust component \citep[$\sim$ 35~K;][]{Xu15}.
In an ISM where the gas and dust are intermingled, the molecular gas component separates into a warm component of $T_{\rm{warm}}$ $\sim$ 110~K and a cold component of $T_{\rm{cold}}$ $\sim$ 35~K.
This can be addressed by RADEX with simple assumptions.

In order to model the warm and cold gas components, we minimize the number of free parameters by fixing the cold gas component using the observed line ratios at the outer region contained within a radius of 2\farcs8 - 3\farcs0.
This region shows a good agreement with a single-phase model with a temperature of $T_{\rm{cold}}$ $\sim$ 19~K (Figure~\ref{figure_6}a).
Here we assume that cold molecular clouds, which dominate the outer region, dominate the cold gas component in the starburst ring.
We calculate the contribution of the warm gas component to the total CO luminosity by using,
\begin{equation}
r = \frac{L'_{{\rm CO(2-1), warm}}}{L'_{{\rm CO(2-1), obs}}},
\end{equation}
where $L'_{{\rm CO(2-1), warm}}$ is the CO~(2--1) luminosity of the warm gas component and $L'_{{\rm CO(2-1), obs}}$ is the observed CO~(2--1) luminosity within the central 1\farcs8 radius.

\begin{figure}
\begin{center}
\includegraphics[width=6cm]{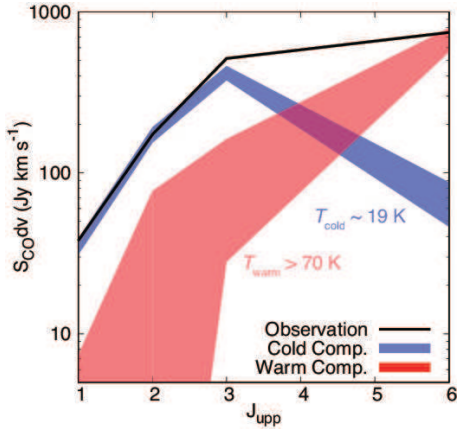}
\caption{Two-phase CO SLED model.  Dark line shows the observed CO SLED toward a central 1\farcs8 (594~pc) radius.  The red and blue areas show the 90\% confidence range of the warm and cold component, respectively.  The areas cover the results from the grid A, B, and C.
}
\label{figure_11}
\end{center}
\end{figure}

In the fitting routine, we varied $r$ within a range of 0 - 0.009 (d$r$ = 0.001), 0.01 - 0.09 (d$r$ = 0.01), and 0.1 - 0.9 (d$r$ = 0.1) to estimate the CO luminosities of the warm gas and cold gas components.
Then, we varied the warm gas kinetic temperature within a range of $T_{\rm{warm}}$ = 10 - 800~K (d$T_{\rm{warm}}$ = 10~K) and the warm gas column density of $N$(H$_2$)$_{\rm{warm}}$ = 10$^{19}$ - 10$^{24}$~cm$^{-2}$ (d$N$(H$_2$)$_{\rm{warm}}$ = 10$^{0.1}$~cm$^{-2}$).
We fixed [CO]/[H$_2$] (= 3 $\times$ 10$^{-4}$) and the gas density of the warm gas component as $n_{\rm{warm}}$ = 10$^{2.0}$ (grid A), 10$^{3.0}$ (grid B), and 10$^{4.0}$~cm$^{-3}$ (grid C).
The results are shown in Figure~\ref{figure_11}.
The minimum $\chi^2$ are 1.30 (grid A), 1.05 (grid B), and 0.85 (grid C), which are smaller than the $\chi^2$ values of the single-phase modeling for the starburst ring (r $<$ 1\farcs8).
The 90\% confidence ranges of $T_{\rm{warm}}$ are 300 - 370, $>$ 250, and 70 - 450~K, whereas those of $N$(H$_2$)$_{\rm{warm}}$ are $>$ 10$^{23.7}$, $>$ 10$^{22.9}$, and 10$^{23.0 - 23.6}$ for the grid A, B, and C, respectively.
The 90\% confidence ranges of $r$ are $<$ 0.1 (best-fit = 0), $<$ 0.8 (best-fit = 0), and $<$ 0.6 (best-fit = 0.003) for the grids A, B, and C, respectively, indicating less contribution of the warm gas component to the lower-$J$ CO fluxes.
This two-phase ISM model in NGC~1614 is similar to that found in the nearby ULIRG Arp~220, showing that a warm gas component has almost no contribution to CO lines lower than $J$ = 3--2 \citep{Rangwala11}.

Using the PDR model (Section~\ref{PDR}) assuming $n$ = 10$^{2.0-5.0}$~cm$^{-3}$, we find that the FUV field ($G_0$) of the inferred warm gas component is higher than 10$^{6.5}$.
Considering the relatively low $\Sigma_{\rm{SFR}}$ \citepalias{Saito15b} and the absence of a heavily obscured AGN, such extreme radiation field is unreasonable for NGC~1614.

In summary, the starburst ring of NGC~1614 can be described by a two-phase molecular gas ISM with a cold component of $\sim$ 19~K (dominated by normal GMCs) and a warm component of $>$ 70~K (warm ISM directly related to powerful activities).
UV-radiation produced by star forming activities (i.e., clumpy PDRs) are not possible heating sources of the warm gas.
We note that these models are consistent with a two-phase dust ISM ($\sim$ 35~K and $\sim$ 110~K) of the starburst ring \citep{Pereira-Santaella15, Xu15} and multi-phase models for other U/LIRGs \citep[e.g.,][]{van_der_Werf10, Rangwala11}.

\subsubsection{Power source of the warm gas}
The observed high-$J$ CO fluxes are systematically higher than those in the Narayanan \& Krumholz model (Section~\ref{nk}).
Also, FUV radiation heating due to the observed star-forming activities cannot reproduce the possible warm gas component (Section~\ref{two}) assuming the two-phase model.
Here we employ the mechanical heating (i.e., shock excitation) as an alternative mechanism to power the warm gas.
We found the largest values of the CO~(2--1)/CO~(1--0) and CO~(3--2)/CO~(1--0) ratios ($\sim$ 10 and $\sim$ 27, respectively) along with the putative outflow as described in Section~\ref{ratio}, which are similar to what has been observed in the nearby Seyfert galaxy NGC~1068 \citep[$\sim$ 10 and $\sim$ 35 respectively at the E-knot;][]{Viti14}.
NGC~1068 has a CND which is thought to be heated by the nuclear jet \citep{Garcia-burillo14}.
The similarity between these galaxies may suggest that in the central 1\farcs8 of NGC~1614 the CO emission is significantly affected by the kinematical interaction between the starburst ring and the putative outflow.
The outflow can heat the starburst ring kinematically, producing highly excited molecular gas components around the shocked region.
This scenario is consistent with the detection of H$_2$ 1--0 S(1) line within the central 1\farcs8 radius \citep{Alonso-Herrero01, Kotilainen01}, suggesting that nearly 1/3 of the total near-IR H$_2$ is produced by shocks.
Stellar feedback such as supernova explosions and stellar winds are also possible sources of mechanical heating.
Therefore, we compare the kinetic luminosity of the outflow and the degree of stellar feedback with the total CO luminosity of the warm gas.

\begin{figure*}
\begin{center}
\includegraphics[width=15cm]{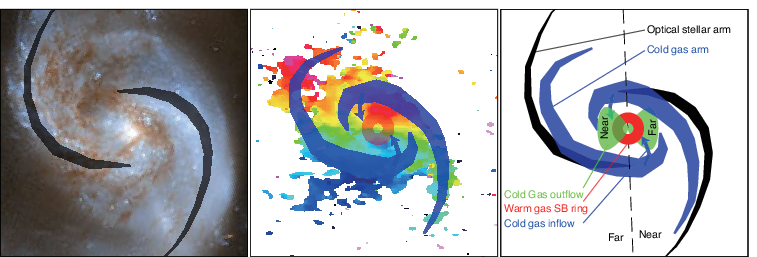}
\caption{(left) Possible optical arms overlaid on the HST/ACS image of NGC~1614 [Credit: NASA, ESA, the Hubble Heritage Team (STScI/AURA)-ESA/Hubble Collaboration and A. Evans (University of Virginia, Charlottesville/NRAO/Stony Brook University)].  (b) Possible main molecular gas arms [blue], cold gas inflow [blue arrows], and the starburst ring detected in CO~(6--5) [red] \citep{Xu15}.  (c) Schematic illustration of the possible geometry of NGC~1614.
}
\label{figure_13}
\end{center}
\end{figure*}

We derive the kinetic luminosity from the supernovae ($L_{{\rm SN}}$) using \citep{Maloney99},
\begin{equation}
L_{{\rm SN}} \sim 3 \times 10^{43} \left(\frac{\nu_{{\rm SN}}}{1\:{\rm yr}^{-1}}\right)\left(\frac{E_{{\rm SN}}}{10^{51}\:{\rm erg}}\right)\:{\rm erg\:s^{-1}},
\end{equation}
where $\nu_{{\rm SN}}$ is the supernova rate and $E_{{\rm SN}}$ is the energy of a supernova \citep[$\sim$ 10$^{51}$ erg;][]{Rangwala11}.
Using the $\nu_{{\rm SN}}$ of $\sim$ 0.9 \citep{Rosenberg12}, we obtain $L_{{\rm SN}}$ of $\sim$ 2.7 $\times$ 10$^{43}$ erg s$^{-1}$.
Assuming the energy from stellar winds is similar to supernovae \citep{McCray&Kafatos87}, the total kinetic luminosity from the stellar feedback is $\sim$ 5.4 $\times$ 10$^{43}$ erg s$^{-1}$.
On the other hand, the kinetic luminosity from the putative outflow is $\sim$ 2.5 $\times$ 10$^{42}$ erg s$^{-1}$ \citep{Garcia-burillo14}.
Comparing with the inferred CO luminosity of the warm gas component ($\sim$ 1.2 $\times$ 10$^{41}$ erg s$^{-1}$; sum up to $J_{\rm{upp}}$ = 23), both the outflow and the stellar feedback are possible heating sources of the warm gas component even when we adopt a low energy injection efficiency of $\lesssim$ 5\%.
Although the molecular outflow is a possible source in terms of the energy, the high resolution CO~(6-5) image \citep{Xu15} shows that the distribution is closely related to the starburst ring, not the outflow (Figure~\ref{figure_12b}a).
This indicates that stellar feedback is more likely to be the heating source for higher-$J$ CO rather than the putative molecular outflow when we consider both energy and spatial CO~(6--5) distribution.
We note that the mechanical energy is also a possible heating source in Arp~220 \citep{Rangwala11}, NGC~6240 \citep{Meijerink13}, and other U/LIRGs \citep{Greve14}.

\subsection{Balance of molecular gas in the starburst ring}
In order to evaluate the effect of the molecular gas inflow, we estimate molecular gas inflow rate (IFR) using,

\begin{equation}\label{eqn5}
{\rm IFR} = \frac{M_{\rm in}v_{\rm in}}{r_{\rm in}}\:M_{\odot}\:{\rm yr}^{-1}
\end{equation}
where $M_{\rm in}$ is the inflowing H$_2$ mass, $v_{\rm in}$ is the inflowing velocity, and $r_{\rm in}$ is the distance from the starburst ring \citep[e.g.,][]{Combes13}.
Assuming the inflowing component is along the gas disk, equation~\ref{eqn5} can be calculated as follows.

\begin{eqnarray}
{\rm IFR} &=& M_{\rm in}\left(\frac{v_{\rm los}}{\cos{\theta}}\right)\left(\frac{\sin{\theta}}{r_{\rm proj}}\right)\nonumber \\
&\sim& 54\left(\frac{\tan{\theta}}{\tan{33\degr}}\right)\left(\frac{\alpha_{\rm CO}}{0.2\:\alpha_{{\rm MW}}}\right)\:M_{\odot}\:{\rm yr}^{-1}
\end{eqnarray}
where $v_{\rm los}$ is the line-of-sight velocity relative to the rotation velocity at the projected distance ($r_{\rm proj}$), $\theta$ is the inclination of the disk, and the $\alpha_{{\rm MW}}$ is the CO-to-H$_2$ conversion factor for the Galaxy \citep[= 4.3;][]{Bolatto13}.
Here we adopt $\theta$ of 33\degr and $\alpha_{{\rm CO}}$ of 0.2$\alpha_{{\rm MW}}$ which are the same values to derive the molecular gas outflow rate \citep[OFR;][]{Garcia-burillo15}.
We estimate the flux density of the inflowing gas components using Figure~\ref{figure_5}e.
The northern (southern) inflowing component has 25.8 (19.3)~Jy km s$^{-1}$, which corresponds to $M_{{\rm in}}$ of 10$^{8.40}\left(\frac{\alpha_{\rm{CO}}}{0.2\:\alpha_{\rm{MW}}}\right)$ (10$^{8.27}\left(\frac{\alpha_{\rm{CO}}}{0.2\:\alpha_{\rm{MW}}}\right)$)~$M_{\odot}$.
The $v_{\rm los}$ and $r_{\rm proj}$ of the northern (southern) component are 80 (40)~km s$^{-1}$ and 500 (500)~pc, respectively.
The derived IFR is 54 $M_{\odot}$ yr$^{-1}$ with the uncertainty by a factor of three, which mainly due to the uncertainties of $v_{\rm los}$ and $r_{\rm proj}$.

Comparing the SFR in the starburst ring \citep[32.8 $M_{\odot}$ yr$^{-1}$;][]{Xu15} and OFR \citep[40 $\pm$ 12 $M_{\odot}$ yr$^{-1}$;][]{Garcia-burillo15}, and assuming that the IFR, OFR, and SFR continue constantly, we can investigate the balance between molecular gas input and output in the starburst ring of NGC~1614.

\begin{equation}
  \frac{{\rm IFR}}{{\rm SFR + OFR}}  \sim \left\{ \begin{array}{ll}
    0.7 & (\theta = 33\degr, \alpha_{{\rm CO}} = 0.2\:\alpha_{{\rm MW}}) \\
    1.2 & (\theta = 33\degr, \alpha_{{\rm CO}} = \alpha_{{\rm MW}})
  \end{array} \right.
\end{equation}
Independent of the adopted $\alpha_{\rm{CO}}$, the molecular gas input and output are roughly balanced, indicating that NGC~1614 can sustain both star-forming activities and an outflow in the starburst ring until exhausting the molecular gas reservoir in the starburst ring and the outer region via bar-driven inflow.
The total molecular gas depletion time due to star formation and outflow (= $M_{\rm{H_2}}$/(SFR + OFR)) ranges from 17.7 $\pm$ 4.6~Myr ($\alpha_{\rm{CO}}$ = $\alpha_{\rm{MW}}$) to 56.6 $\pm$ 9.3~Myr ($\alpha_{\rm{CO}}$ = 0.2$\alpha_{\rm{MW}}$).
This is consistent with the typical duration timescale of U/LIRG phase predicted by numerical merger simulations \citep[dozens of Myr;][]{Saitoh09, Teyssier10}.
The luminous merger remnant NGC~1614 may quench itself until the next dozens of Myr due to depletion of the cold molecular gas reservoir (if the molecular outflow and star formation continue constantly).

\subsection{Central Gas Geometry of NGC~1614} \label{illust}
We summarize all results provided by this Paper and previous studies into a simple schematic picture (Figure~\ref{figure_13}).
We used a position angle of $\sim$ -352\degr and an inclination of $\sim $36\degr as estimated by \citet{Ueda14}, for the cold gas disk and the starburst ring.
This is consistent with the results provided by \citet{Garcia-burillo15}.
We found ``S"-shape non-circular motions along the leading edge of the north-south near-IR bar in the lower-$J$ CO velocity fields and PVDs.
The near-IR bar connects to the near-IR spiral arms \citep{Olsson10}.
Moreover, the northern part of the molecular bar was detected in SMA CO~(2--1) observations \citep[dashed line in Figure~3 of][]{Konig13}.
\citet{Konig16} also suggested that the northern molecular gas components are diffuse gas clouds which are now funneled into the nuclear region, although the flowing direction is different from the picture shown in Figure~\ref{figure_13}.
The central starburst ring \citep[SFR $\sim$ 32.8 $M_{\rm{\odot}}$ yr$^{-1}$;][]{Xu15} was detected in star formation tracers \citep{Alonso-Herrero01, Herrero-Illana14, Saito15b, Pereira-Santaella15} and dense/warm gas tracers \citep{Sliwa14, Xu15}.
The nucleus has no strong AGN \citep{Imanishi&Nakanishi13, Herrero-Illana14, Pereira-Santaella15} or heavily obscured AGN \citep{Xu15}.
The upper limit of the AGN luminosity suggests that it might be possible for driving the putative molecular gas outflow with the outflow rate of 40 $M_{\rm{\odot}}$ yr$^{-1}$ as well as the starburst activities \citep{Garcia-burillo15}.
The detection of the molecular outflow coincides with the geometry of the ionized gas outflow with no redshifted component due to the extinction by the large column of dust in the inclined disk \citep{Bellocchi12}.

Radiative transfer modeling with RADEX suggest that the starburst ring will have a single-phase ($\sim$ 42~K) or two-phase ($\sim$ 19~K and $>$  70~K) molecular gas ISM, while the outer disk has a single cold gas ISM ($\sim$ 22~K).
The single-phase model is consistent with the FUV radiation field estimated by PDR models.
However, star-forming activities are not enough to power the warm gas component in the starburst ring if the two-phase ISM is valid.
Alternatively, we suggest the mechanical heating from the supernovae and stellar winds as a heating source of the warm gas component.

\section{Conclusion} \label{conclusion}
In this Paper, we present a detailed study of the molecular gas ISM in the luminous merger remnant NGC~1614 through high-resolution, high-sensitivity, and $uv$-matched ALMA observations of the CO~(1--0), CO~(2--1), CO~(3--2), and CO~(6--5) lines.
The results are summarized as follows:

\begin{enumerate}
\item The CO~(6--5) line shows a compact distribution which coincides with the starburst ring detected in Pa $\alpha$ and radio-to-FIR continuum emission, while the other lower-$J$ CO lines are extended.
This indicates that the CO~(6--5) can be used as a better tracer of star formation as already suggested by \citet{Xu15}.
\item We find a ``S"-shape non-circular motion along the near-IR north-south bar in the CO~(1--0), CO~(2--1), and CO~(3--2) velocity fields.
Comparing with numerical simulations, this may be a bar-driven cold gas inflow connecting to the starburst ring.
\item Radiative transfer modeling with RADEX reveal that the molecular gas ISM in NGC~1614 can be described in a single-phase or two-phase model.
The single-phase model shows a radial gradient of the gas kinetic temperature from 70 to 20~K and the H$_2$ column density from 10$^{22.9}$ to 10$^{22.0}$ cm$^{-2}$ with peaks at the starburst ring.
This single-phase model is applicable to explain the observed CO SLEDs, while there is a systematic underestimation of the CO~(3--2) and CO~(6--5) flux.
The Narayanan \& Krumholz CO SLED model using the observed $\Sigma_{\rm{SFR}}$ also shows the same trend.
To account for the underestimation in both models, we verify the two-phase model to the line ratios in the central 1\farcs8 radius.
\item The two-phase RADEX model for the starburst ring shows a good agreement with a cold gas component of $\sim$ 19~K and a warm gas component of $>$ 70~K, while there are number of assumptions at this stage.
This is consistent with the two-phase dust ISM ($\sim$ 35~K and $\sim$ 110~K) revealed by high-resolution ($<$ 0\farcs5) mid-IR and far-IR observations and two-phase modelings for other nearby U/LIRGs.
Higher-$J$ CO observation ($J_{\rm{upp}}$ $\geq$ 7) is critical to characterize the molecular gas ISM in the starburst ring of NGC~1614.
\item Considering that the observed star-forming activities are not enough to power the warm gas component inferred from the two-phase RADEX, we suggest mechanical heating from supernovae and stellar winds as an alternative power source.
\item The summation between the cold molecular gas outflow rate (OFR) and the star formation rate (SFR) in the starburst ring is comparable to the cold gas inflow rate (IFR), showing evidence of the balance between the molecular gas input and output at the starburst ring.
Assuming IFR, OFR, and SFR continue constantly, it takes dozens of Myr to exhaust all H$_2$ mass in NGC~1614.
This is consistent with the typical U/LIRG duration timescale suggested by numerical merger simulations.

\end{enumerate}

We will present our high-resolution multi-transition observations, including CO,  $^{13}$CO, CN, CS, etc., toward NGC~1614 using Band~3 and Band~6 of ALMA in a forthcoming paper (Ando et al. in preparation).

\acknowledgements

The authors thank an anonymous referee for coments that improved the contents of this paper.
TS thanks N. Lu, E. W. Pellegrini, and E. Schinnerer for useful discussion.
TS and the other authors thank ALMA staff for their kind support and H. Nagai for the instruction of ALMA data reduction.
TS and ML are financially supported by a Research Fellowship from the Japan Society for the Promotion of Science for Young Scientists.
TS was supported by the ALMA Japan Research Grant of NAOJ Chile Observatory, NAOJ-ALMA-0114.
DI was supported by the ALMA Japan Research Grant of NAOJ Chile Observatory, NAOJ-ALMA-0011, and JSPS KAKENHI Grant Number 15H02074.
This paper makes use of the following ALMA data: ADS/JAO.ALMA\#2011.0.00182.S, ADS/JAO.ALMA\#2011.0.00768.S, ADS/JAO.ALMA\#2013.1.00991.S, and ADS/JAO.ALMA\#2013.1.01172.S.
ALMA is a partnership of ESO (representing its member states), NSF (USA) and NINS (Japan), together with NRC (Canada), NSC and ASIAA (Taiwan), and KASI (Republic of Korea), in cooperation with the Republic of Chile.
The Joint ALMA Observatory is operated by ESO, AUI/NRAO and NAOJ.
This research has made extensive use of the NASA/IPAC Extragalactic Database (NED) which is operated by the Jet Propulsion Laboratory, California Institute of Technology, under contract with the National Aeronautics and Space Administration.


\begin{deluxetable*}{lcccccccccrl}
\tabletypesize{\scriptsize}
\tablecaption{Log of ALMA Observations \label{table_obs}}
\tablewidth{0pt}
\tablehead{
Band & UT date &\multicolumn{4}{c}{Configuration} &$T_{\rm{sys}}$ &\multicolumn{3}{c}{Calibrator} &$T_{\rm{integ}}$ \\
\cline{3-6} \cline{8-10}
 & &Array &FoV & $N_{\rm{ant}}$ &$L_{\rm{baseline}}$ & &Flux &Bandpass &Gain & \\
 & & & (\arcsec) & & (m) & (K) & & & & (min.) \\
(1) &(2) &(3) &(4) &(5) &(6) &(7) &(8) &(9) &(10) &(11)
 }
\startdata
B3 &2014 Jun. 30 &ACA 7~m &95.6 &10 &7.3 - \phantom{00}48 &\phantom{0}42 - \phantom{0}170 &J0501-0159 &J0423-013 &J0427-0700 &49.1 \\
B3 &2014 Dec. 04 &ALMA 12~m &55.7 &31 &\phantom{.}15 - \phantom{0}290 &\phantom{0}31 - \phantom{0}120 &Uranus &J0423-0120 &J0423-0120 &13.1 \\
B3\footnotemark &2014 Aug. 30 &ALMA 12~m &55.7 &35 &\phantom{.}28 - 1060 &\phantom{0}50 - \phantom{0}130 &J0423-013 &J0423-0120 &J0423-0120 &16.9 \\
B6\footnotemark[a] &2014 Aug. 14 &ALMA 12~m &27.8 &35 &\phantom{.}15 - \phantom{0}349 &\phantom{0}70 - \phantom{0}160 &Uranus &J0423-0120 &J0423-0120 &7.3 $\times$ 3 \\
B7 &2012 Jul. 31/Aug. 14 &ALMA 12~m &18.6 &27 &\phantom{.}18 - \phantom{0}431 &100 - \phantom{0}210 &Callisto &J0522-364 &J0423-013 &88.8 \\
B9 &2012 Aug. 13/28 &ALMA 12~m &\phantom{0}9.3 &23 &\phantom{.}20 - \phantom{0}394 &500 - 1100 &Ceres &J0423-013 &J0423-013 &49.4
\enddata
\tablecomments{
Column 2: Observed date.
Column 3: Dish size.
Column 4: FWHM of the primary beam at the frequency of the target CO line.
Column 5: Number of available antennas.
Column 6: Projected length of assigned baseline for NGC~1614.
Column 7: DSB system temperature toward NGC~1614.
Column 8--10: Observed flux calibrator, bandpass calibrator, and phase calibrator.
Column 11: Total integration time on NGC~1614.  The Band~6 observation has three pointing fields.
}
\tablenotetext{a}{Data observed for our ALMA Cycle~2 program (ID: 2013.1.01172.S)}
\end{deluxetable*}

\begin{deluxetable*}{cccccccccccc}
\tabletypesize{\scriptsize}
\tablecaption{CO Line and Imaging Properties\label{table_data}}
\tablewidth{0pt}
\tablehead{
CO Line  & $\nu_{\rm{rest}}$ & $E_{\rm{u}}/k$ &Band &MRS &$uv$-weight &Beam size &rms & $S_{\rm{CO}}\Delta v$ &Recovered flux & Ref. \\
 & (GHz) & (K) & & (\arcsec) & & (\arcsec) & (mJy b$^{-1}$) & (Jy km s$^{-1}$) & (\%) & \\
 (1) &(2) &(3) &(4) &(5) &(6) &(7) &(8) &(9) & (10) & (11)
 }
\startdata
$J$ = 1--0 (ACA+12~m) &115.271 &\phantom{00}5.53 &B3 &17.6 &briggs &1.0 $\times$ 0.6 &1.4 &295.3 $\pm$ 1.8 &113 $\pm$ 26 &1, a \\
$J$ = 1--0 (12~m) &115.271 &\phantom{00}5.53 &B3 &\phantom{0}4.6 &briggs &1.0 $\times$ 1.0 &1.6 &\phantom{0}69.3 $\pm$ 1.1 &\phantom{0}27 $\pm$ \phantom{0}5 &This work, a \\
$J$ = 2--1 (12~m) &230.542 &\phantom{0}16.60 &B6 &\phantom{0}4.6 &uniform &1.0 $\times$ 1.0 &3.5 &325.4 $\pm$ 2.7 &\phantom{0}40 $\pm$ \phantom{0}9 &This work, b \\
$J$ = 3--2 (12~m) &345.796 &\phantom{0}33.19 &B7 &\phantom{0}4.6 &briggs &1.0 $\times$ 1.0 &3.0 &866.8 $\pm$ 2.2 &\phantom{0}59 $\pm$ \phantom{0}6 &2, c \\
$J$ = 6--5 (12~m) &691.473 &116.16 &B9 &\phantom{0}4.6 &briggs &1.0 $\times$ 1.0 &8.0 &826.9 $\pm$ 3.2 &\phantom{0}58 $\pm$ 10 &2, 3, d
\enddata
\tablecomments{
Column 1: Observed CO transition.
Column 2: Rest frequency of the transition.
Column 3: Upper state energy of the transition.
Column 5: Maximum recoverable scale (MRS) of the visibility data; See text.
Column 6: Visibility ($uv$-plane) weighting for imaging. ``briggs" means Briggs weighting with {\tt robust} = 0.5.
Column 7: Convolved beam size.  All the images, except for the ACA-combined CO~(1--0) image, in this paper are convolved into the same resolution of 1\farcs0, which is larger than the synthesized beam size.
Column 8: Noise rms for data with the velocity resolution of 30 km s$^{-1}$.
Column 9: Integrated intensity inside the 3$\%$ contour of the peak value.  We only consider the statistical error in this column.
Column 10: The ALMA flux divided by the single dish flux. We consider the statistical and the systematic error of the flux calibration.
Column 11: Reference of ALMA data (1. \citet{Konig16}, 2. \citet{Sliwa14}, 3. \citet{Xu15}) and single dish data (a. NRAO~12m; \citet{Sanders91}, b. SEST~15m; \citet{Albrecht07}, c. JCMT~15m; \citet{Wilson08}, d. Herschel/SPIRE; \citet{Xu15}).
}
\end{deluxetable*}

\appendix
\section{Channel maps of the CO SLED of NGC~1614} \label{A1}
\begin{figure*}
\begin{center}
\includegraphics[width=15cm]{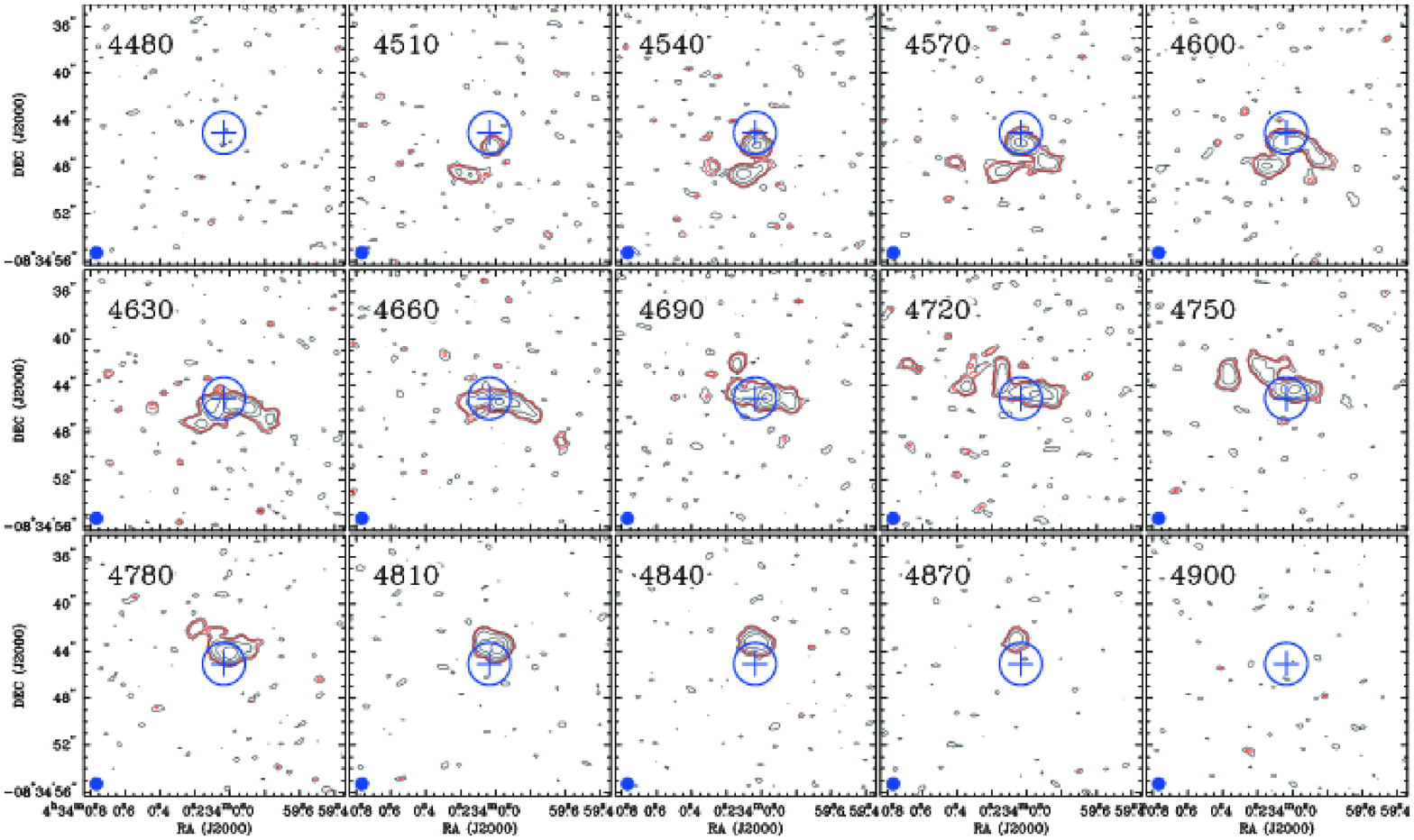}
\includegraphics[width=15cm]{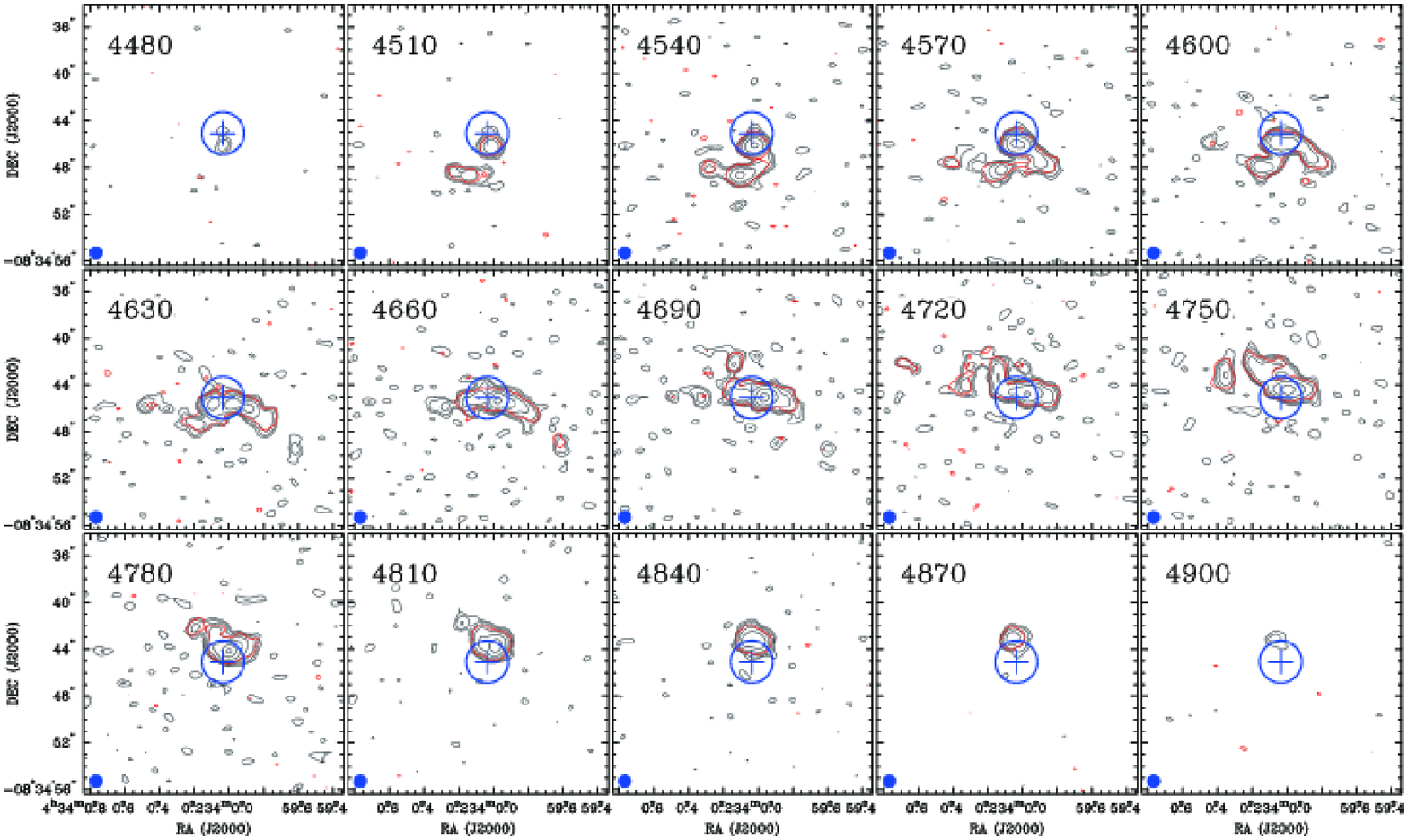}
\caption{Channel maps (black contours) of CO~(1--0) [top] and CO~(2--1) [bottom] in NGC~1614.  The $n$th contours are at 2$^{n}\sigma$ ($n$ = 1, 2, 3 ...).  $\sigma$ is the noise rms listed in Table~\ref{table_data}.  The red contours show the 4$\sigma$
 level of the CO~(1--0) as a comparison.  The blue cross indicates the nucleus which is detected in Pa $\alpha$ and the radio continuum emission \citep{Olsson10,Herrero-Illana14}.  The blue circle indicates the approximate outer edge of the nuclear ring (\citealt{Herrero-Illana14, Xu15}; \citetalias{Saito15b}).}
\label{app_cube1}
\end{center}
\end{figure*}

\begin{figure*}
\begin{center}
\includegraphics[width=15cm]{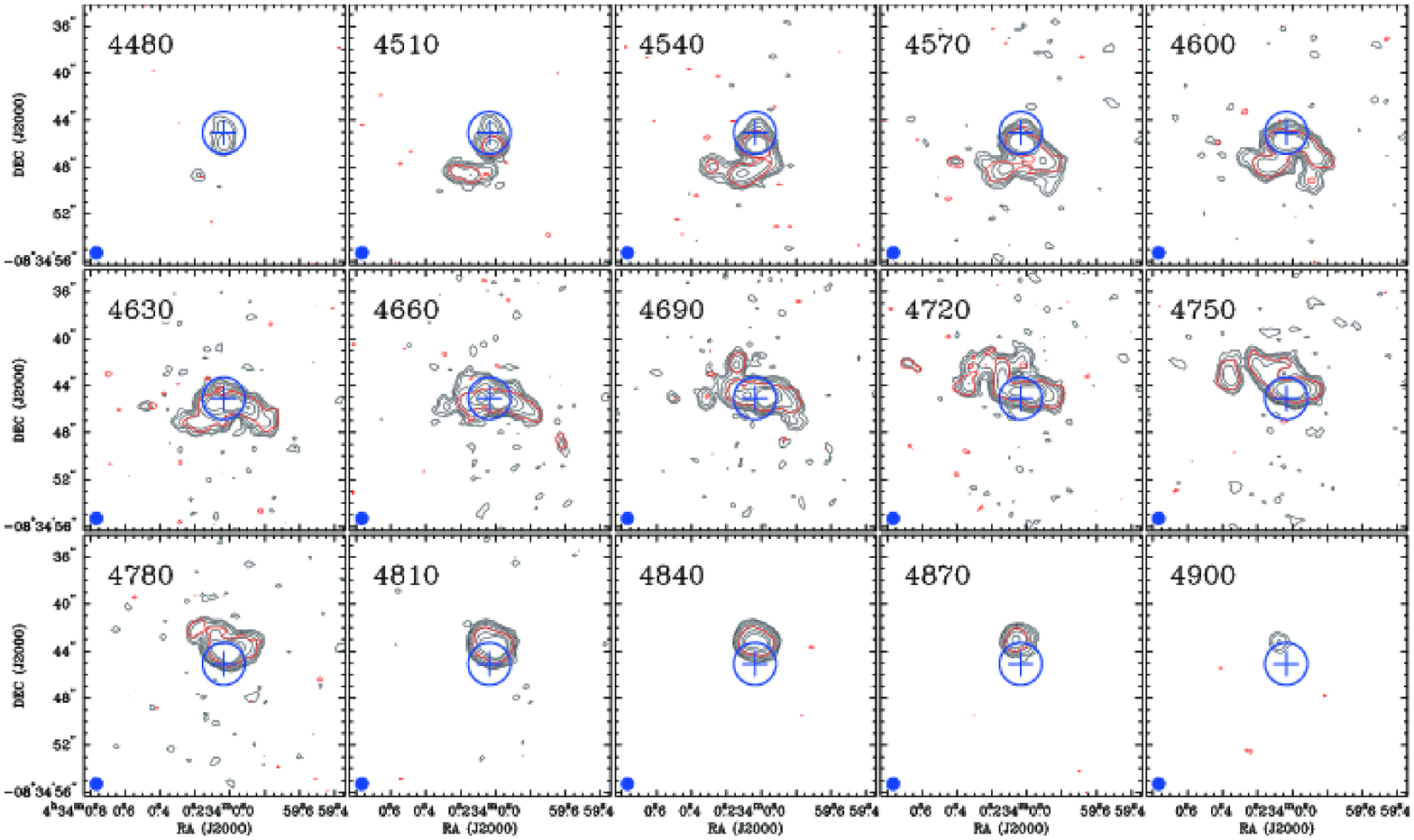}
\includegraphics[width=15cm]{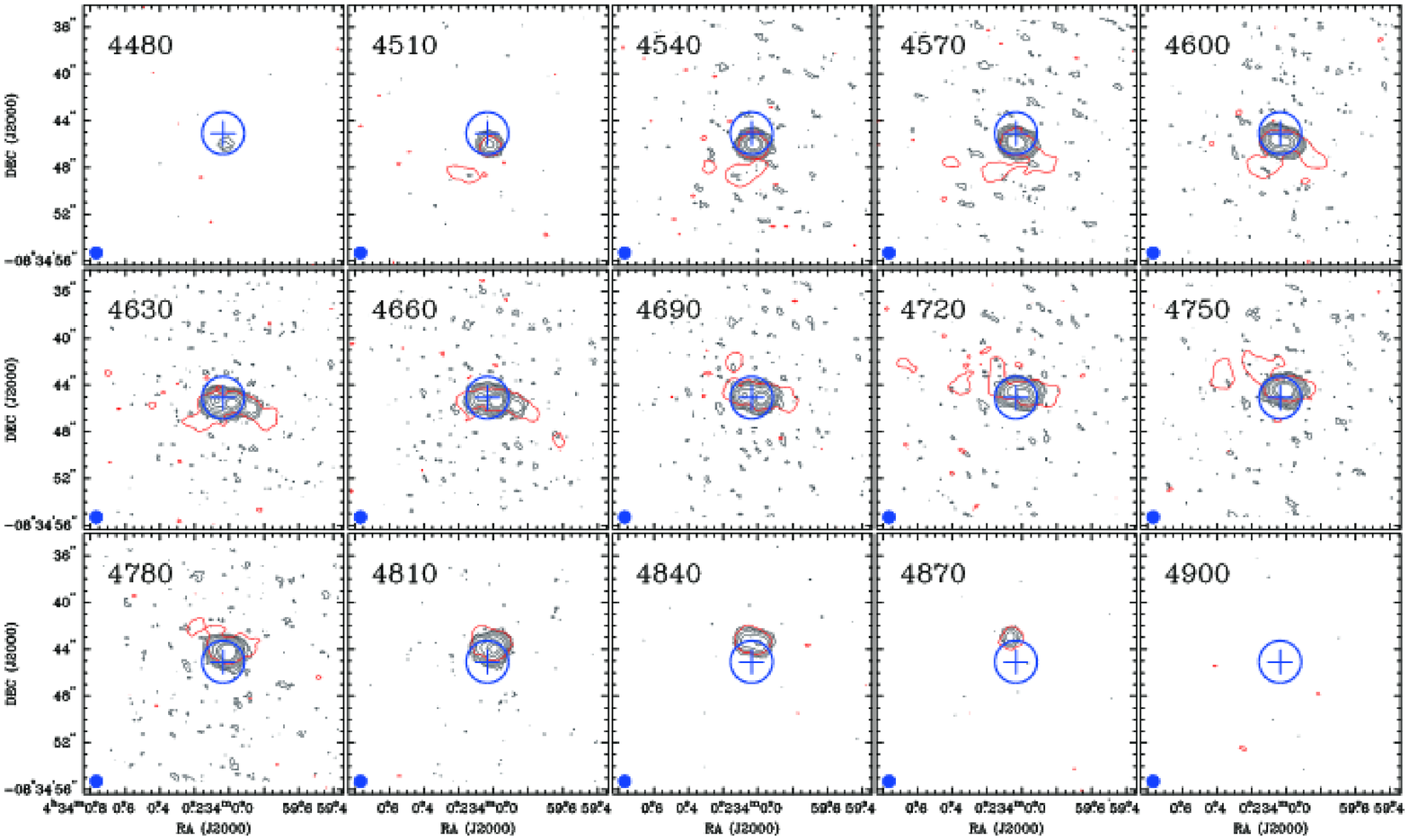}
\caption{Same as Figure~\ref{app_cube1} but for CO~(3--2) (top) and CO~(6--5) (bottom).
}
\label{app_cube2}
\end{center}
\end{figure*}

\begin{figure*}
\begin{center}
\includegraphics[width=15cm]{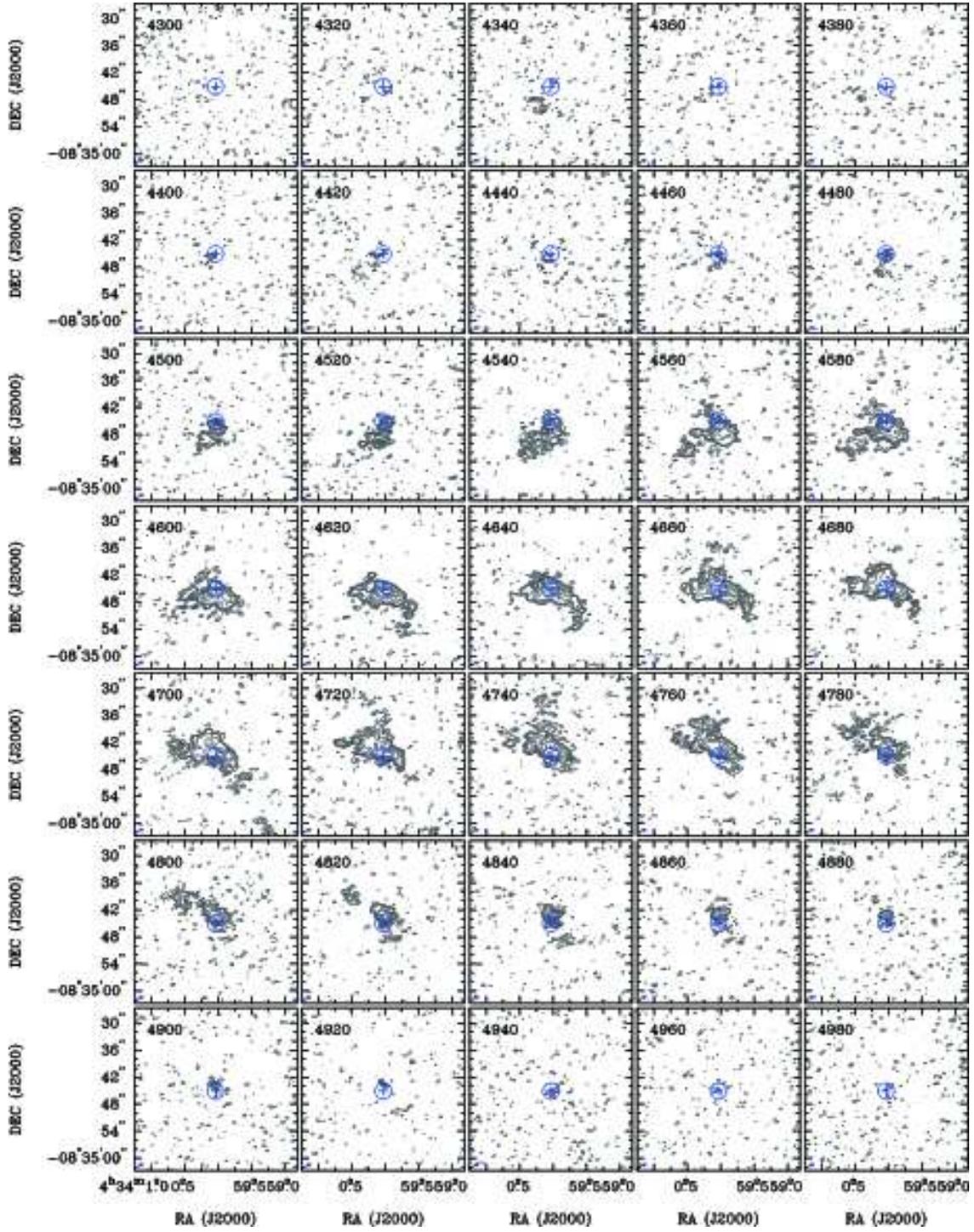}
\caption{Same as Figure~\ref{app_cube1} but for the ACA-combined CO~(1--0).  The contours are 1$\sigma$ $\times$ (-2, 2, 3, 4, 8, 16, and 32) mJy beam$^{-1}$.
}
\label{app_cube3}
\end{center}
\end{figure*}

We show the channel map of the $uv$-matched CO~(1--0), CO~(2--1), CO~(3--2), and CO~(6--5), and ACA-combined CO~(1--0) in Figure~\ref{app_cube1}, \ref{app_cube2}, and \ref{app_cube3}.
These maps have the same angular resolution of 1\farcs0 $\times$ 1\farcs0, the same velocity resolution of 30~km s$^{-1}$, and the same MRS of 45~k$\lambda$ ($\simeq$ 4\farcs6), except for the ACA-combined CO~(1--0) image.
The lower-$J$ three transitions show almost same structures although the CO~(6--5) is only restricted within the central 2\farcs3 radius.

\clearpage

\end{document}